\documentclass[aps,prl,twocolumn,superscriptaddress,showpacs]{revtex4-1}
\usepackage{amssymb,amsmath,amsfonts,graphicx,epsf}
\usepackage{color}

\begin{document}

\title{Kinetic turbulence in relativistic plasma: from thermal bath to non-thermal continuum}

\author{Vladimir Zhdankin}
\email{zhdankin@jila.colorado.edu}
\affiliation{JILA, University of Colorado and NIST, 440 UCB, Boulder, CO 80309}
\author{Gregory R. Werner}
\affiliation{Center for Integrated Plasma Studies, Physics Department, University of Colorado, 390 UCB, Boulder, CO 80309}
\author{Dmitri A. Uzdensky}
\affiliation{Center for Integrated Plasma Studies, Physics Department, University of Colorado, 390 UCB, Boulder, CO 80309} 
\affiliation{Institute for Advanced Study, Princeton, NJ 08540} 
\author{Mitchell C. Begelman}
\affiliation{JILA, University of Colorado and NIST, 440 UCB, Boulder, CO 80309}
\affiliation{Department of Astrophysical and Planetary Sciences, 391 UCB, Boulder, CO 80309}

\date{\today}

\begin{abstract}
We present results from particle-in-cell simulations of driven turbulence in magnetized, collisionless, and relativistic pair plasma. We find that fluctuations are consistent with the classical $k_\perp^{-5/3}$ magnetic energy spectrum at fluid scales and a steeper $k_\perp^{-4}$ spectrum at sub-Larmor scales, where $k_\perp$ is the wavevector perpendicular to the mean field. We demonstrate the development of a non-thermal, power-law particle energy distribution, $f(E) \sim E^{-\alpha}$, with index $\alpha$ that decreases with increasing magnetization and increases with increasing system size (relative to the characteristic Larmor radius). Our simulations indicate that turbulence can be a viable source of energetic particles in high-energy astrophysical systems, such as pulsar wind nebulae, if scalings asymptotically become insensitive to the system size.
\end{abstract}

\pacs{52.27.Ep, 52.27.Ny, 52.35.Ra, 52.65.Rr}

\maketitle

{\em Introduction.}--- 

Non-thermal energetic particles are a common ingredient in high-energy astrophysics, being responsible for observable broadband radiation spectra. However, their origin is often poorly constrained. It is now widely accepted that non-thermal particle acceleration can be a consequence of collisionless plasma physics. In particular, it was demonstrated that relativistic magnetic reconnection \citep{sironi_spitkovsky_2014, guo_etal_2014, werner_etal_2015} and collisionless shocks \citep{spitkovsky_2008, sironi_spitkovsky_2010} can both efficiently energize a population of non-thermal particles with a power-law energy distribution. Since these two mechanisms require specific large-scale configurations to initiate, however, it is unclear whether they are sufficiently versatile to explain all observations.

A third route to particle acceleration, which may be both versatile and ubiquitous, is turbulence in collisionless plasmas. Turbulence is often inevitable in astrophysical flows due to the large separation between driving scales and dissipative scales, making it an attractive possible source of energetic particles. In particular, turbulence may power non-thermal synchrotron and inverse Compton radiation in systems such as pulsar wind nebulae, coronae of accreting objects, and jets. Furthermore, intermittent dissipative structures may naturally explain impulsive flares observed in some of these systems, including GeV flares in the Crab nebula \citep{abdo_etal_2011, tavani_etal_2011}.

Turbulent particle acceleration can be associated with a number of acceleration mechanisms, which are not necessarily independent from those in shocks and magnetic reconnection. Indeed, large-scale turbulence may intermittently accelerate particles through first-order Fermi acceleration in self-consistently formed shocks and reconnection sites \citep[e.g.,][]{stone_etal_1998, servidio_etal_2009, eyink_etal_2013, zhdankin_etal_2013}, while small-scale, instability-driven turbulence is essential for Fermi acceleration in shocks and reconnection \citep{fermi_1949, blandford_eichler_1987}. Particles can also be stochastically accelerated via wave-particle interactions, generally leading to second-order Fermi acceleration \citep[e.g.,][]{cho_lazarian_2006, chandran_etal_2010, petrosian_2012, lazarian_etal_2012, beresnyak_li_2016}. However, it remains largely unestablished under what circumstances the turbulent cascade can efficiently accelerate particles toward a robust non-thermal particle energy distribution.

In this Letter, we utilize particle-in-cell (PIC) simulations to demonstrate that kinetic turbulence in collisionless, relativistically-hot pair plasmas can efficiently generate a non-thermal particle population from an initial thermal bath in closed systems of modest size. The simulations are driven to develop a classical large-scale magnetohydrodynamic (MHD) cascade that transitions into a kinetic cascade at sub-Larmor scales. The late-time particle distributions take the form of power laws that span a broad range of energies. For a fixed system size, these power-law distributions become harder with increasing magnetization (or decreasing plasma beta) and qualitatively resemble those previously seen in relativistic magnetic reconnection \citep{guo_etal_2014, guo_etal_2015, werner_etal_2015}. However, the distributions become steeper with increasing system size, indicating that asymptotic, system-size independent scalings either have not yet been reached or do not exist.

{~}\\
{\em Method.}--- 

Hydrodynamic and MHD simulations show that relativistic turbulence broadly resembles the non-relativistic case \citep[][]{cho_2005, zrake_macfadyen_2011, zrake_macfadyen_2012, radice_rezzolla_2013, cho_lazarian_2013, zrake_east_2016}, but can only describe acceleration in the test particle approximation \citep[e.g.,][]{kowal_etal_2012, lynn_etal_2014}. In this work, we apply kinetic PIC simulations, which, for hot pair plasmas, can obtain a comparable inertial range \citep[][]{makwana_etal_2015} while self-consistently describing particle acceleration.  Previously, PIC simulations were applied to show the emergence of non-thermal features from decaying turbulence \citep[e.g.,][]{nalewajko_etal_2016, yuan_etal_2016, makwana_etal_2016_arxiv} and from the magnetorotational instability \citep{hoshino_2013, riquelme_etal_2012, hoshino_2015}.

In our system, the magnetic field $\boldsymbol{B}(\boldsymbol{x},t)$ and electric field $\boldsymbol{E}(\boldsymbol{x},t)$ are evolved by Maxwell's equations, while the electron/positron particles are evolved via the Lorentz force. The characteristic kinetic scales are the Larmor radius $\rho_e \equiv \bar{\gamma} m c^2 / e B_{\rm rms}$ and plasma skin depth $d_e \equiv \sqrt{\bar{\gamma} m c^2 / 4 \pi n_0 e^2}$, given mean particle Lorentz factor~$\bar{\gamma} \gg 1$, electron/positron rest mass~$m$, elementary charge~$e$, speed of light~$c$, mean total particle density~$n_0$, and characteristic (rms) magnetic field~$B_{\rm rms}$. The two free dimensionless parameters (ignoring driving parameters) are the system size relative to Larmor radius~$L / \rho_e$ and nominal magnetization $\sigma \equiv B_{\rm rms}^2 / 4 \pi n_0 \bar{\gamma} m c^2 = (d_e/\rho_e)^2$ (inversely proportional to plasma beta). We denote initial values by $\sigma_0 \equiv \sigma(t=0)$ and $\rho_{e0} \equiv \rho_e(t=0)$.

We performed our simulations using the explicit electromagnetic PIC code Zeltron \citep{cerutti_etal_2013}. The domain is a periodic cubic box of size $L^3$ with uniform background magnetic field $\boldsymbol{B}_0 = B_0 \hat{\boldsymbol{z}}$. We initialize simulations with zero electromagnetic fluctuations ($\delta \boldsymbol{B} = \boldsymbol{E} = 0$) and particles with uniform non-drifting Maxwell-J\"{u}ttner distribution \citep{synge_1957} and ultra-relativistic mean Lorentz factor~$\bar{\gamma}(t=0) \approx 300$. This stable thermal equilibrium is disrupted by external driving. To drive strong, critically-balanced turbulence at large scales in a way that mimics energy transfer from an MHD cascade  \citep{goldreich_sridhar_1995}, we apply a fluctuating external current density $\boldsymbol{J}_{\rm ext}$ in the form of an oscillating Langevin antenna \citep{tenbarge_etal_2014}. We drive $J_{{\rm ext},z}$ at eight modes, $\boldsymbol{k}_0 L / 2 \pi \in \{ (1,0,\pm1), (0,1,\pm1), (-1,0,\pm1), (0,-1,\pm1) \}$, and each of $J_{{\rm ext},x}$ and $J_{{\rm ext},y}$ in four modes to enforce $\nabla \cdot \boldsymbol{J}_{\rm ext} = 0$. We choose driving frequency $\omega_0 = 0.6 \cdot 2 \pi v_{A0} / \sqrt{3} L$ and decorrelation rate $\Gamma_0 = 0.5 \cdot 2 \pi v_{A0} / \sqrt{3} L$, where $v_{A0} = c \sqrt{\sigma_0/(\sigma_0 + 4/3)}$ is the initial relativistic Alfv\'{e}n velocity in our simulations \citep{sakai_kawata_1980, gedalin_1993}. We tune the driving amplitude such that rms fluctuations satisfy $\delta B_{\rm rms} \sim B_0$.

Since there is no energy sink in our numerical set-up, injected energy will increase fluid internal energy linearly in time at a heating rate (per unit volume) comparable to $\epsilon \sim B_0^2 c/8 \pi L$. This heating will cause~$\rho_e(t)$ to increase in time and~$\sigma(t)$ to decrease in time, with the dimensionless parameter $\xi \equiv \sigma \rho_e/L$ being statistically constant in time. The parameter $\xi$ can be expressed as $\xi = {\cal E}_{\rm mag}/{\cal E}_{\rm max}$, where ${\cal E}_{\rm mag} = B_{\rm rms}^2/8\pi n_0$ is magnetic energy per particle and ${\cal E}_{\rm max} = L e B_{\rm rms} / 2 c$ is the maximum energy of particles for given system size (i.e., with Larmor radius equal to half the system size). All simulations with fixed $\xi$ but varying $\sigma$ nominally represent different stages of evolution for a single run. Fully-developed turbulence begins after a few light crossing times (once the cascade reaches kinetic scales) and ends when the fluid inertial range is suppressed by heating ($\rho_e \sim L/2\pi$). A rough estimate for the duration of turbulence, assuming $\rho_e(t) \sim \rho_{e0} + t \epsilon / n_0 e B_{\rm rms}$, is $tc/L \sim L / 2 \pi \rho_{e0} \sigma_0 \sim 1 / 2 \pi \xi(t=0)$.

{~}\\
{\em Results.}--- 

\begin{figure}
\includegraphics[width=\columnwidth]{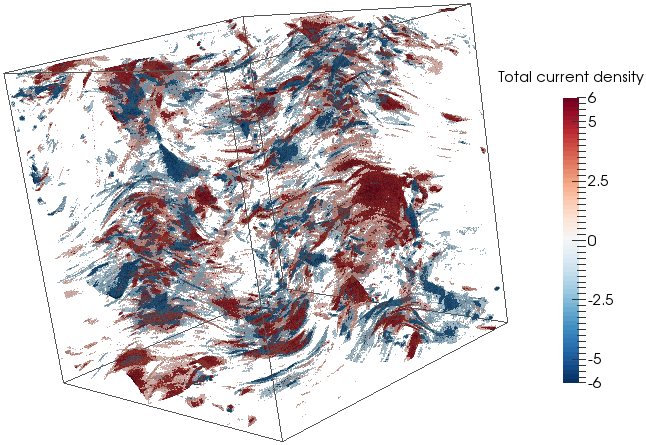} 
\includegraphics[width=0.8\columnwidth]{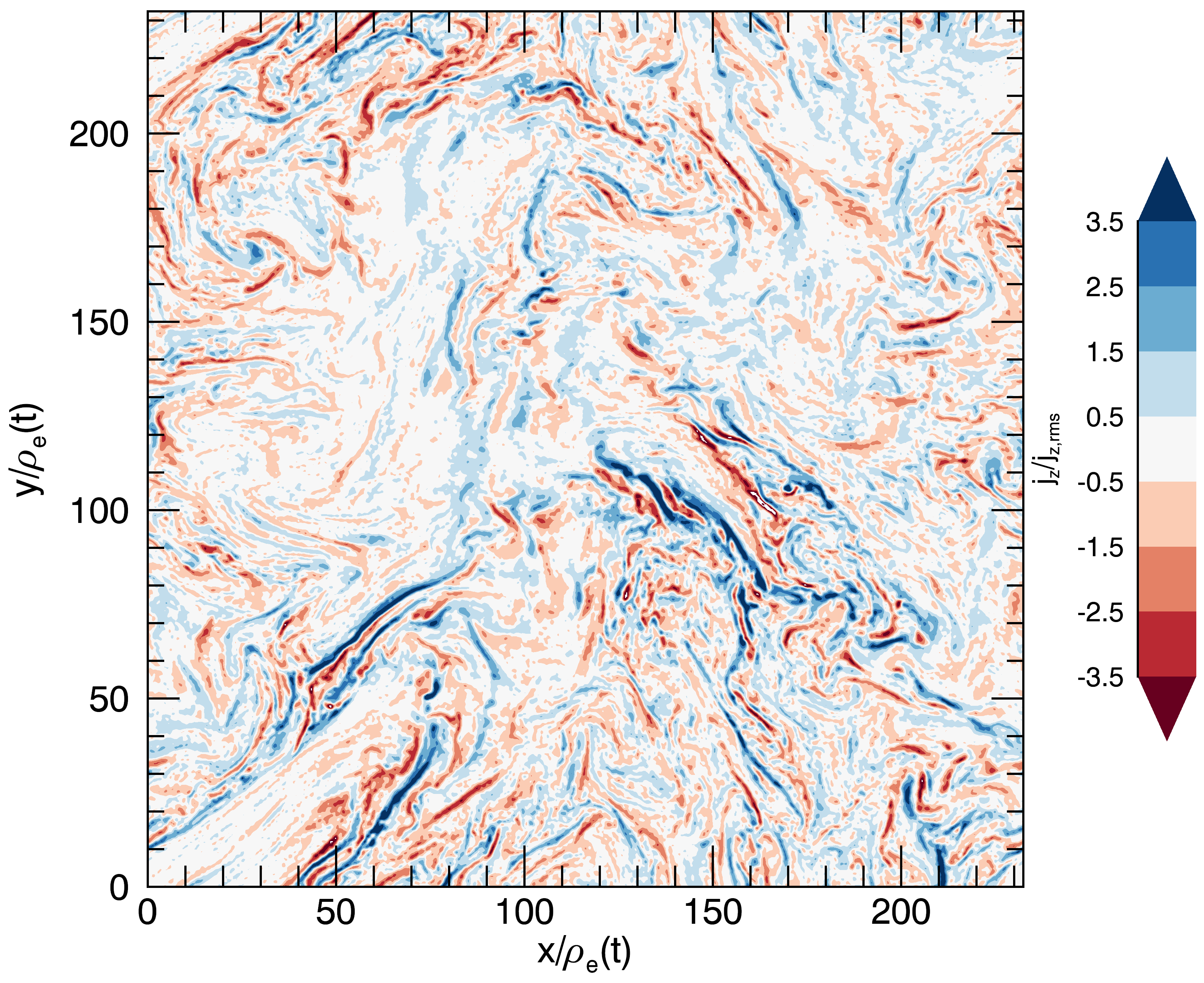}
   \centering
   \caption{\label{fig:profile} 3D and 2D threshold plots of $J_z/J_{z,\rm rms}$ (normalized electric current density parallel to $\boldsymbol{B}_0$).}
 \end{figure}

We performed a series of simulations on lattices of varying size~$N^3$ with varying parameters~$\sigma_0$ and~$L/\rho_{e0}$. For simulations with $N \in \{256,384,512,768,1024\}$, we chose a corresponding ratio of driving scale to initial Larmor radius no greater than $L/2\pi\rho_{e0} \in \{108.6,81.5,54.3,40.7,27.2\}$. For simulations with~$N \le 512$, we performed a full scan of $\sigma_0 \in \{0.25,0.5,1,2,4\}$; for~$N = 768$ we did $\sigma_0 \in \{0.25,0.5,1,2\}$; and for~$N = 1024$ we did $\sigma_0 \in \{0.5,2\}$. Unless otherwise noted, we describe results from our fiducial $768^3$ simulation with $\sigma_0 = 0.25$ and $L/2\pi\rho_{e0} \approx 61.1$ (with $\sim 10^{11}$ particles). For all runs, we chose $\rho_{e0} \ge 1.5 \Delta x$ (where $\Delta x$ is the lattice cell size), at least $128$ particles per cell\footnote{Particle distributions are converged when $128$ or more particles per cell are used; see Supplementary Material at [URL will be inserted by publisher]}, and a duration $\ge 10 L/c$ (including $\sim 22 L/c$ for the fiducial case); note that the Alfv\'{e}n crossing time is longer than $L/c$ and slowly increases in time. In most of our cases, total energy (accounting for injection) is conserved to approximately $1\%$ or better. We show a snapshot of $J_z/J_{z,\rm rms}$, the normalized electric current density along $\boldsymbol{B}_0$, in Fig.~\ref{fig:profile}. The formation of intermittent current sheets, with thicknesses near the kinetic scale and lengths spanning a range of scales up to the driving scale, is evident, as previously seen in MHD turbulence \citep[e.g.,][]{zhdankin_etal_2016b} and in non-relativistic kinetic turbulence \citep[e.g.,][]{makwana_etal_2015, wan_etal_2016}.

\begin{figure}
\includegraphics[width=\columnwidth]{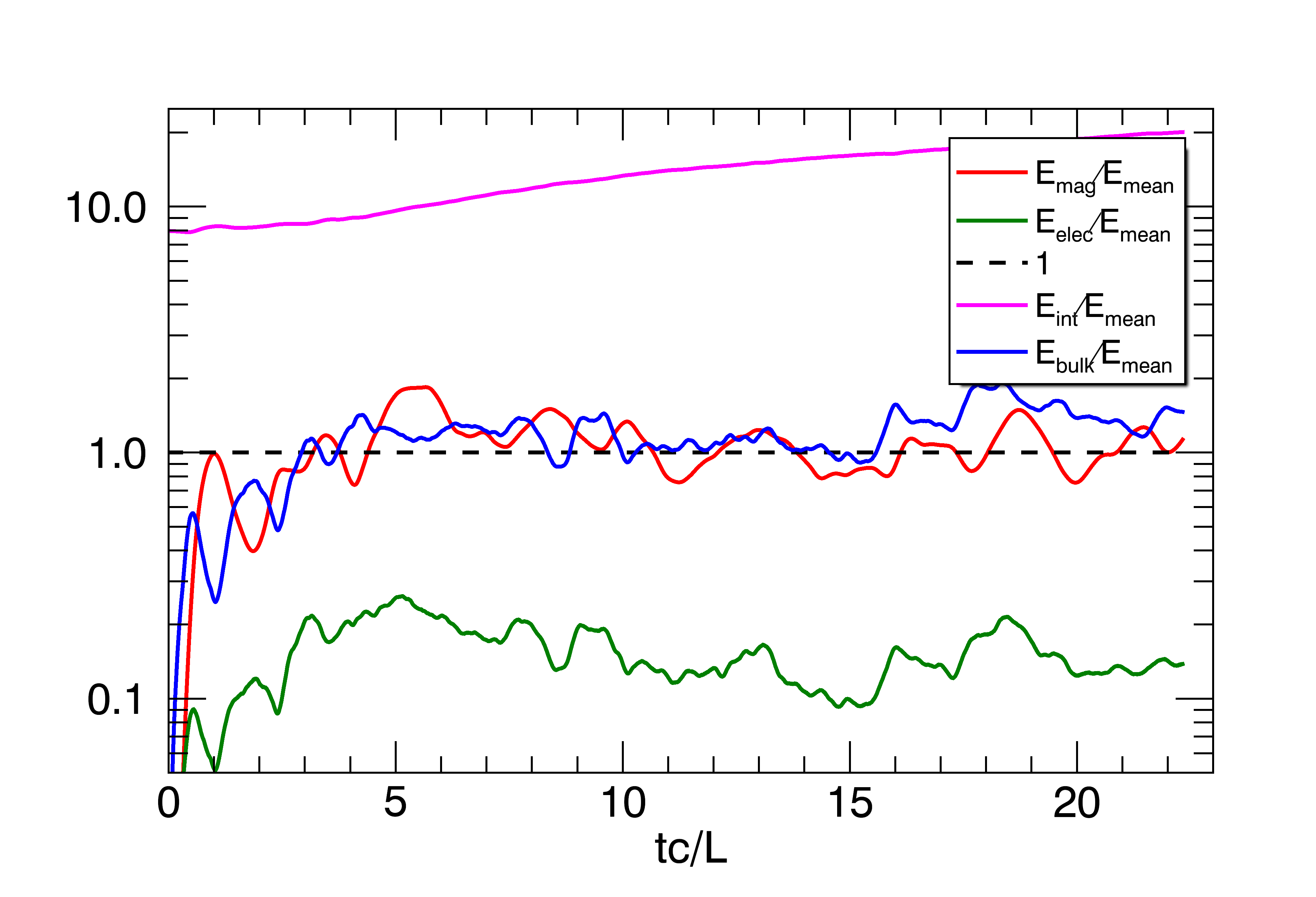}
      \centering
   \caption{\label{fig:evolution} Evolution of turbulent magnetic energy $E_{\rm mag}$ (red), electric energy $E_{\rm elec}$ (green), internal energy $E_{\rm int}$ (magenta), and bulk fluid energy $E_{\rm bulk}$ (blue), all normalized to background magnetic energy $E_{\rm mean}$ (black, dashed).}
 \end{figure}

We first show the evolution of the energy contributions in Fig.~\ref{fig:evolution}. We decompose total particle kinetic energy~$E_{\rm kin}(t) = \int d^3x E_f(\boldsymbol{x},t)$ into internal fluid energy~$E_{\rm int}(t) = \int d^3x [E^2_f(\boldsymbol{x},t) - p^2_f(\boldsymbol{x},t) c^2]^{1/2}$ and bulk fluid kinetic energy~$E_{\rm bulk}(t) = E_{\rm kin}(t) - E_{\rm int}(t)$, where $E_f(\boldsymbol{x},t) = \int d^3p (m^2 c^4 + p^2 c^2)^{1/2} f (\boldsymbol{p}, \boldsymbol{x},t)$ is fluid energy density,~$\boldsymbol{p}_f(\boldsymbol{x},t) = \int d^3p \boldsymbol{p} f (\boldsymbol{p},\boldsymbol{x},t)$ is fluid momentum density, and~$f(\boldsymbol{p},\boldsymbol{x},t)$ is the particle distribution function at time~$t$. Bulk fluid energy can also be expressed as~$E_{\rm bulk}(t) = \int d^3x w^2(\boldsymbol{x},t)$, where~$\boldsymbol{w} \equiv \boldsymbol{p}_f c / [E_f + (E_f^2 - p_f^2 c^2)^{1/2}]^{1/2}$ is fluid four-velocity weighted by an effective mass. We find that turbulent energy saturates after a few light crossing times. The turbulent magnetic energy~$E_{\rm mag}(t) = \int d^3x [\delta \boldsymbol{B}(\boldsymbol{x},t)]^2/8 \pi$ and $E_{\rm bulk}$ come into equipartition with background magnetic energy $E_{\rm mean} = \int d^3x B_0^2/8\pi$, as dictated by our driving, while electric energy~$E_{\rm elec}(t) = \int d^3x [\boldsymbol{E}(\boldsymbol{x},t)]^2/8 \pi$ is a few times smaller. For $\sigma_0 \leq 1$, turbulence energies are significantly below $E_{\rm int}$, which sets fluid inertia, so turbulent motions are effectively non-relativistic. To a good approximation, internal energy increases linearly in time, as expected from a constant energy injection rate. For the fiducial case, $\rho_e$ and~$d_e$ increase by less than a factor of two over the given duration.

We next consider Fourier power spectra for turbulent fluctuations. For simplicity, we consider the magnetic energy spectrum $E_{\rm mag}(\boldsymbol{k}) = |\tilde{\boldsymbol{B}}(\boldsymbol{k})|^2/8\pi$, electric energy spectrum $E_{\rm elec}(\boldsymbol{k}) = |\tilde{\boldsymbol{E}}(\boldsymbol{k})|^2/8\pi$, solenoidal fluid energy spectrum $E_{\rm bulk,sol}(\boldsymbol{k}) = |\hat{\boldsymbol{k}} \times \tilde{\boldsymbol{w}}(\boldsymbol{k})|^2$, and compressive fluid energy spectrum $E_{\rm bulk,comp}(\boldsymbol{k}) = |\hat{\boldsymbol{k}} \cdot \tilde{\boldsymbol{w}}(\boldsymbol{k})|^2$, where $\tilde{\boldsymbol{y}}(\boldsymbol{k})$ denotes the Fourier transform of $\boldsymbol{y}(\boldsymbol{x})$. The decomposition of fluid energy into solenoidal and compressive parts is customary in compressible turbulence \citep[e.g.,][]{kritsuk_etal_2007, federrath_etal_2010}. Each spectrum integrates into the corresponding contribution to total energy.

\begin{figure}
\includegraphics[width=\columnwidth]{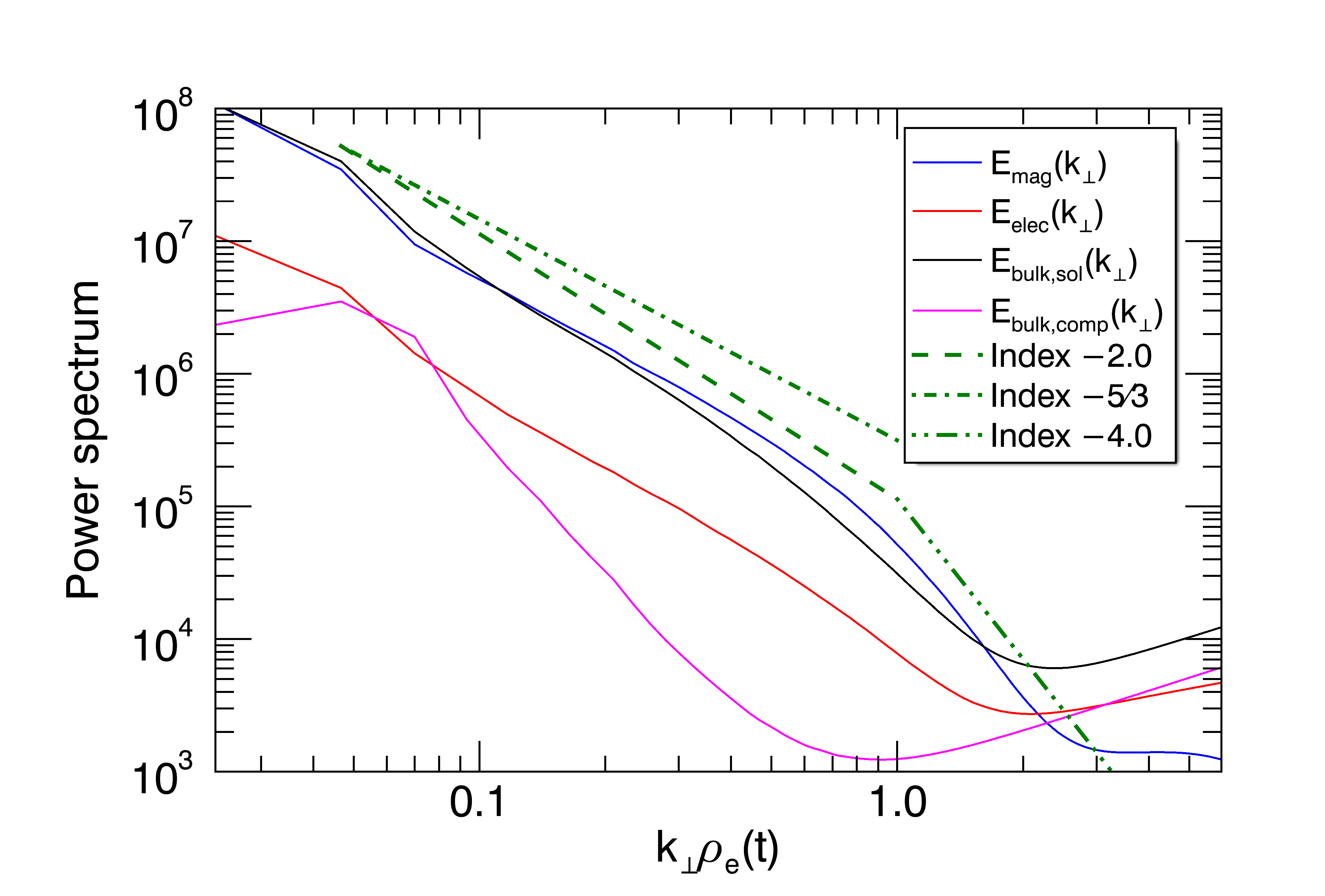}
\includegraphics[width=\columnwidth]{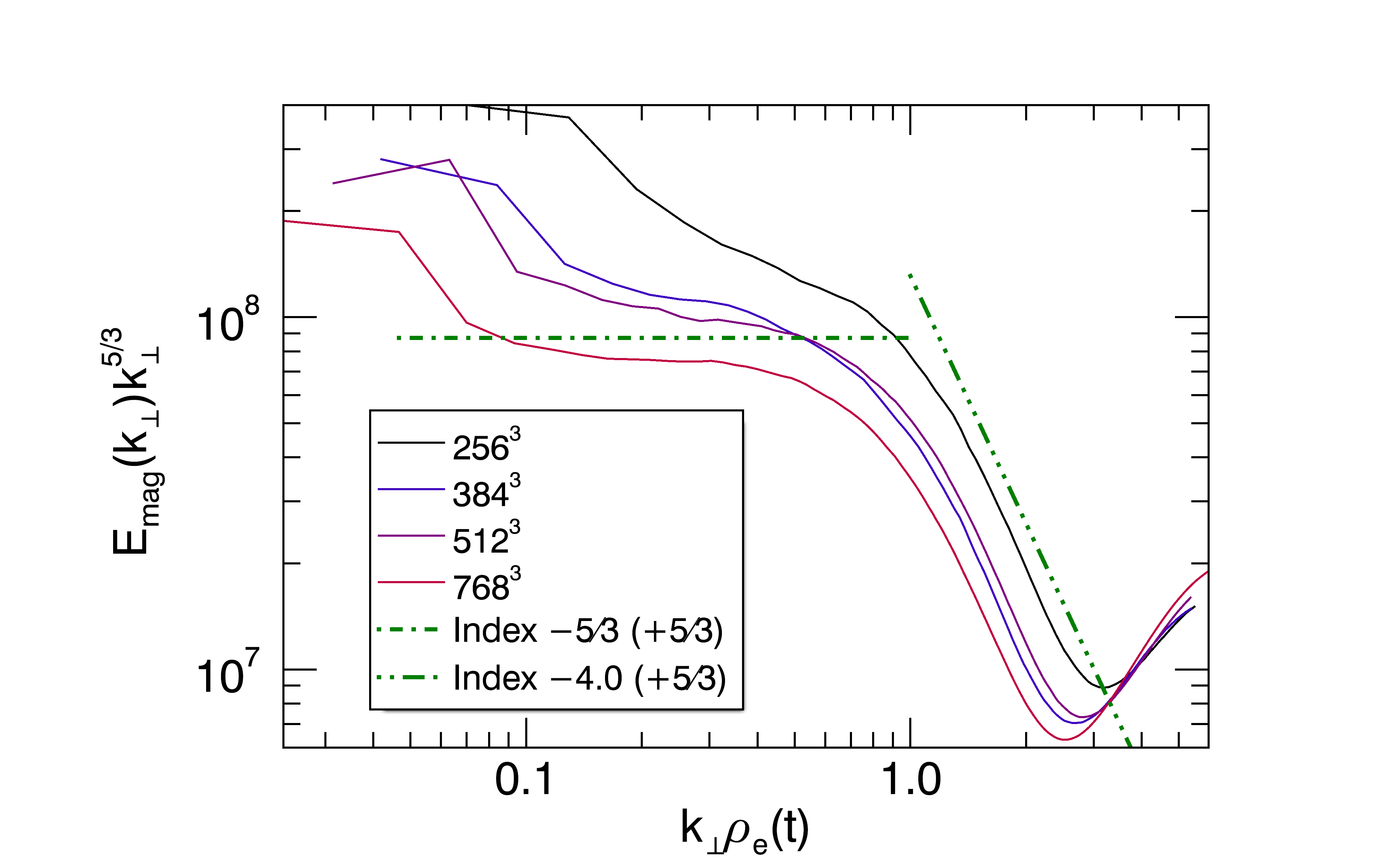}
   \centering
   \caption{\label{fig:spectrum} Top panel: Spectra of magnetic energy $E_{\rm mag}(k_\perp)$ (blue), electric energy $E_{\rm elec}(k_\perp)$ (red), solenoidal fluid kinetic energy $E_{\rm bulk,sol}(k_\perp)$ (black), and compressive fluid kinetic energy $E_{\rm bulk,comp}(k_\perp)$ (magenta) during fully developed turbulence. Green dashed lines indicate $k_\perp^{-5/3}$ and $k_\perp^{-2}$ in the inertial range ($k_\perp \rho_e < 1$) and $k_\perp^{-4}$ in the sub-Larmor range ($k_\perp \rho_e > 1$). Bottom panel: compensated magnetic energy spectrum $E_{\rm mag}(k_\perp) k_\perp^{5/3}$ for simulations of varying system size $L/2\pi\rho_{e0} \in \{ 27.2, 40.7, 54.3, 81.5 \}$ and fixed $\sigma_0 = 0.25$, with $k_\perp^{-5/3}$ and $k_\perp^{-4}$ scalings (green, dashed).}
 \end{figure}

In Fig.~\ref{fig:spectrum}, we show (ring-averaged) spectra~$E_{\rm mag}(k_\perp)$, $E_{\rm elec}(k_\perp)$, $E_{\rm bulk,sol}(k_\perp)$, and~$E_{\rm bulk,comp}(k_\perp)$ for wavevectors~$k_\perp$ perpendicular to $\boldsymbol{B}_0$, averaged over times $11 L/c \lesssim t \lesssim 22 L/c$. We find that magnetic energy and solenoidal fluid energy are in excellent equipartition across the fluid inertial range, with $E_{\rm mag}(k_\perp)$ having a scaling close to $k_\perp^{-5/3}$ while $E_{\rm bulk,sol}(k_\perp)$ has a somewhat steeper scaling closer to $k_\perp^{-2}$. The electric energy and compressive fluid energy are subdominant, as expected in the sub-relativistic, weakly compressible turbulence regime; in particular, $E_{\rm bulk,comp}(k_\perp)$ decreases very rapidly with $k_\perp$ (i.e., steeper than $k_\perp^{-3}$). There is a spectral break for~$E_{\rm mag}(k_\perp)$ at~$k_\perp \rho_e \sim 1$, in the vicinity of which there is an excess of magnetic energy over fluid energy, possibly due to energy exchange associated with kinetic instabilities of anisotropic, non-thermal particle distributions. Beyond the spectral break, $E_{\rm mag}(k_\perp)$ steepens into a power law $k_\perp^{-4}$, implying that the cascade may continue as a kinetic cascade \citep{schekochihin_etal_2009, schoeffler_etal_2014}. At even higher $k_\perp$, spectra flatten due to particle noise. To better characterize the inertial range, we show the compensated magnetic energy spectrum $E_{\rm mag}(k_\perp) k_\perp^{5/3}$ in the second panel of Fig.~\ref{fig:spectrum} and compare to simulations of smaller $L/\rho_{e0}$ (and fixed $\sigma_0 = 0.25$). The magnetic energy spectrum approaches a scaling consistent with $k_\perp^{-5/3}$ for increasing $L/\rho_{e0}$, as predicted for incompressible MHD turbulence \citep{goldreich_sridhar_1995} and for highly relativistic MHD turbulence \citep{thompson_blaes_1998}; kinetic energy spectra steeper than $k_\perp^{-5/3}$ are anticipated in both compressive \citep{federrath_2013} and relativistic \citep{zrake_macfadyen_2012} turbulence.

\begin{figure}
\includegraphics[width=\columnwidth]{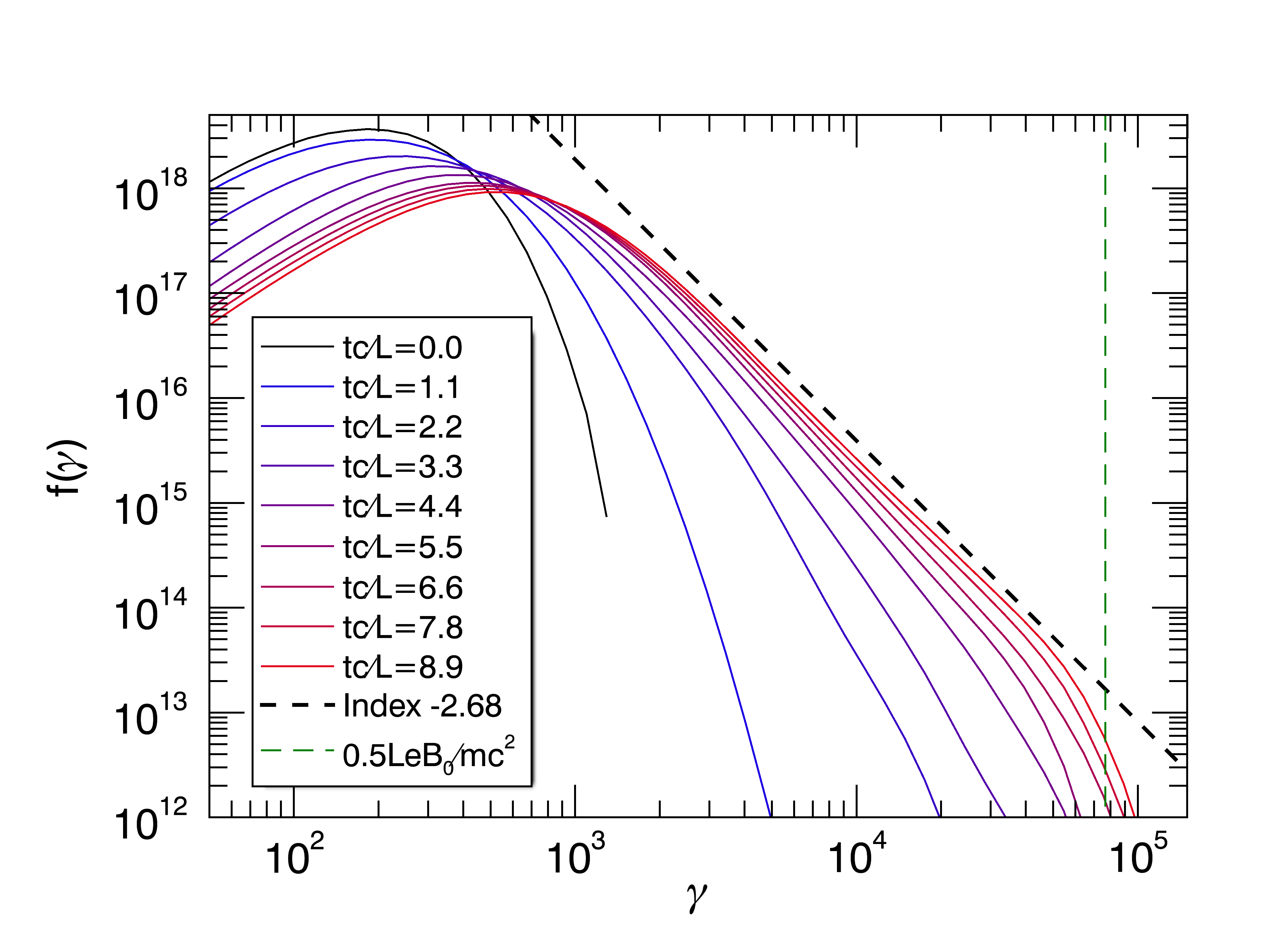}
\includegraphics[width=\columnwidth]{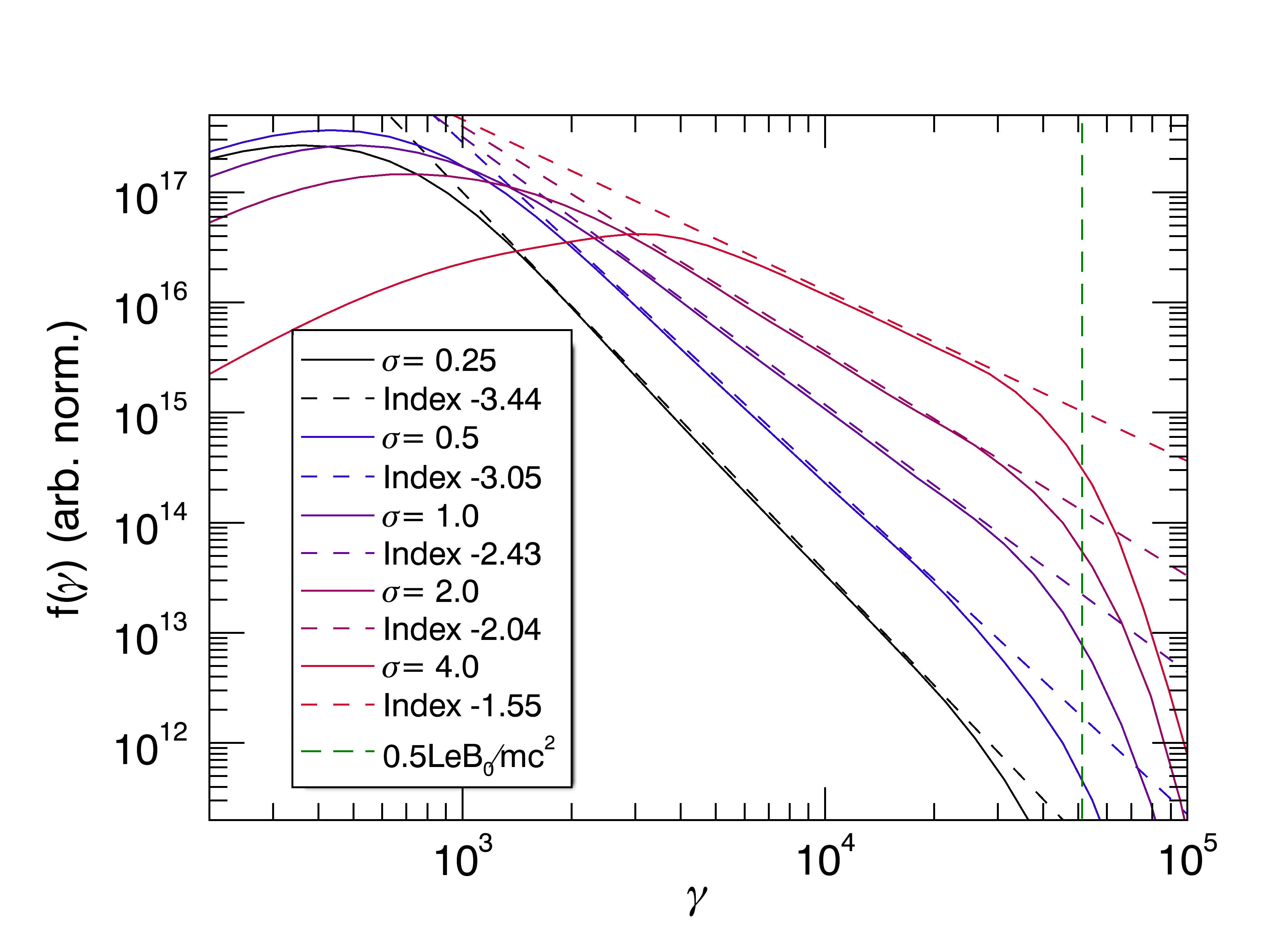}
\includegraphics[width=\columnwidth]{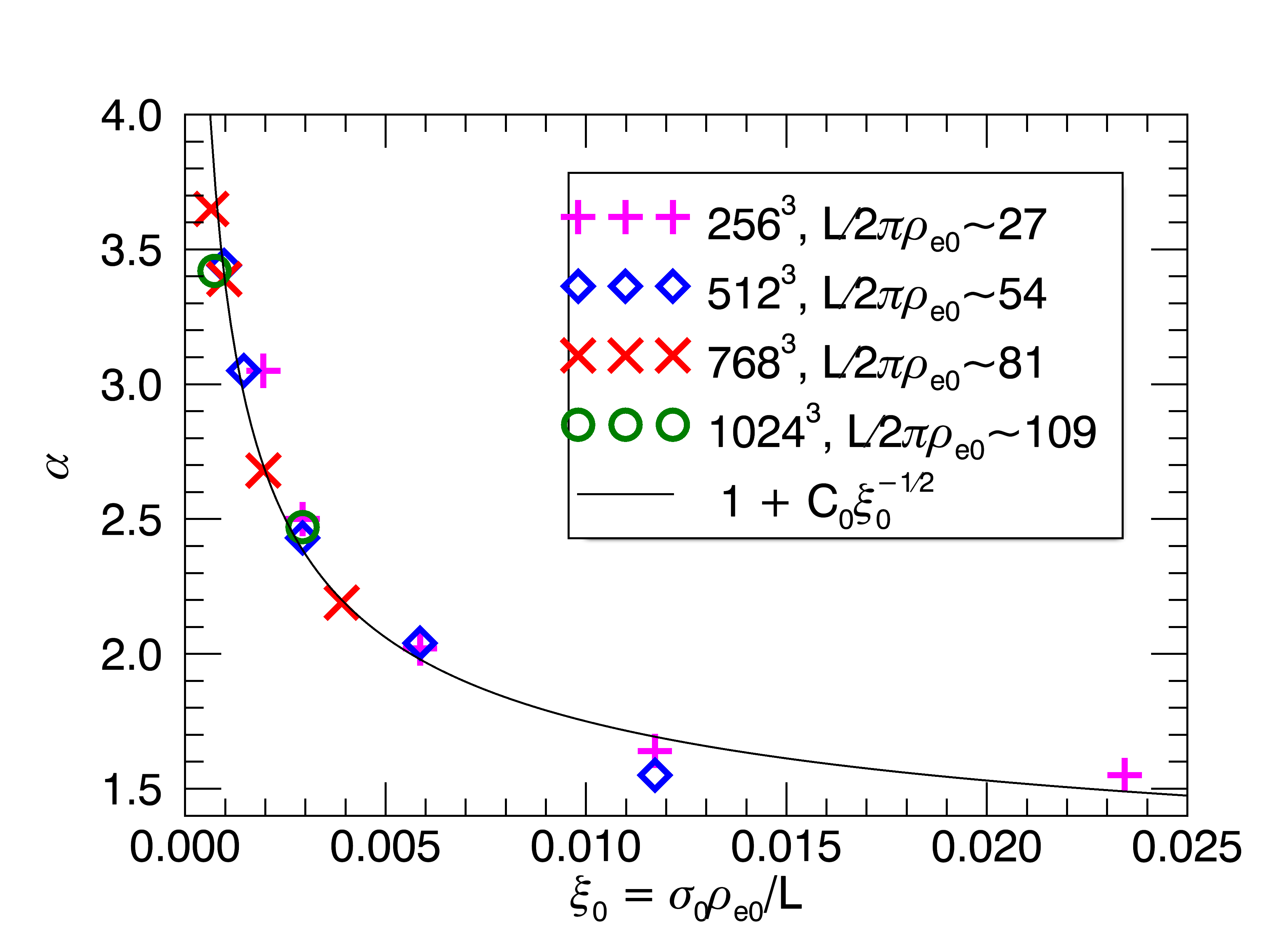}
      \centering
   \caption{\label{fig:dist} Top panel: Evolution of particle energy distribution $f(\gamma)$ (where $\gamma = E/mc^2$) from initial thermal distribution (black) to saturated non-thermal distribution at $t = 8.9 L/c$ (red) (for $768^3$ lattice, $\sigma_0 = 1$), along with system-size cutoff $\gamma_{\rm max}$ (green, dashed) and power law $p^{-2.68}$ (black, dashed). Middle panel: late-time $f(\gamma)$ for $\sigma_0 \in \{1/4,1/2,1,2,4\}$ ($512^3$ lattice; renormalized for clarity), with respective power-law fits $p^{-\alpha}$, $\alpha \in \{3.44, 3.05, 2.43, 2.04, 1.55\}$ (dashed lines). Bottom panel: measured $\alpha$ versus $\xi_0 = \sigma_0 \rho_{e0}/L$, with fit $1 + C_0 \xi_0^{-1/2}$ (black) given $C_0 = 0.075$. Data points are from $256^3$ (magenta, plus), $512^3$ (blue, diamond), $768^3$ (red, cross), and $1024^3$ (green, circle) simulations with nominal $L/2\pi\rho_{e0} \in \{ 27, 54, 81, 109 \}$, respectively.}
 \end{figure}

We now turn to our main result: the production of a non-thermal population of energetic particles by turbulence. We demonstrate in Fig.~\ref{fig:dist} that the particle momentum distribution converges to a power law after several dynamical times ($t \gtrsim 5 L/v_{A0}$): $f(\gamma) \sim \gamma^{-\alpha}$, where $\gamma = E / m c^2$ is the particle Lorentz factor. In the second panel of Fig.~\ref{fig:dist}, we show that $\alpha$ decreases with increasing $\sigma_0$ for fixed system size, and $\alpha < 2$ for sufficiently large $\sigma_0$. A qualitatively similar dependence of $\alpha$ on $\sigma$ was previously found in relativistic magnetic reconnection \citep{guo_etal_2014, guo_etal_2015, werner_etal_2015}. In all of our cases, the upper cutoff for the power law is set by system size, i.e., $\gamma_{\rm max} = L e B_0 / 2 m c^2$. Since particles are unable to significantly exceed $\gamma_{\rm max}$, a bump forms near $\gamma_{\rm max}$ at late times, but does not strongly influence the power law. The lower cutoff grows on the heating timescale and therefore shortens the power law for high $\sigma_0$. In addition to the $\sigma_0$ dependence, we find that the power-law distributions become steeper with increasing $L/\rho_{e0}$. We show the measured values of $\alpha$ for all of our simulations, each taken from the time with the longest fitted power-law segment, versus $\xi_0 = \sigma_0 \rho_{e0}/L$ in the third panel of Fig.~\ref{fig:dist}. We find that the index can be estimated, in all of our simulations, by the empirical formula $\alpha \sim 1 + C_0 \xi_0^{-1/2}$, where $C_0 \approx 0.075$.
 
A preliminary investigation of particle acceleration mechanisms via particle tracking indicates that the acceleration process is diffusive in momentum space, consistent with second-order Fermi acceleration. Predicting the late-time power-law index $\alpha$ analytically is complicated, however, by the time dependence of physical parameters and absence of a cooling mechanism or particle escape.

{~}\\
{\em Conclusions.}---

In this Letter, we demonstrated efficient particle acceleration by driven turbulence in magnetized, collisionless, and relativistically hot plasmas for modestly large, closed domains. Our PIC simulations successfully reproduce large-scale MHD turbulence, as implied by the $k_\perp^{-5/3}$ magnetic energy spectrum. The $k_\perp^{-4}$ spectrum at sub-Larmor scales may indicate a kinetic cascade \citep{schekochihin_etal_2009}, and is also purportedly measured in the solar wind at scales below the electron gyroscale \citep{sahraoui_etal_2009, sahraoui_etal_2013} (although contested \citep[e.g.,][]{alexandrova_etal_2009, alexandrova_etal_2012}). The late-time particle energy distributions are robust power laws that become shallower with increasing values of $\xi = \sigma \rho_e/L$. In particular, we proposed an empirical formula for the power-law index, $\alpha \sim 1 + C_0 \xi^{-1/2}$, which fits all of our simulations to a good approximation. For sufficiently high magnetization, we find that $\alpha < 2$, implying that non-thermal particles are energetically dominant and therefore kinetic physics cannot be neglected. This study provides a first step towards building a self-consistent, first-principles physical picture of driven kinetic turbulence in relativistic collisionless plasmas. 

For turbulence to be a viable source of energetic particles in high-energy astrophysical systems such as pulsar wind nebulae, the late-time power-law index~$\alpha$ must asymptotically become insensitive to the system size~$L$ (relative to $\rho_e$). There are no clear signs of $\alpha$ converging with $L$ in our simulations, and so, if extrapolated to astrophysical system sizes ($L \gg \sigma \rho_e$ so that $\xi \ll 1$), our empirical formula would rule out efficient turbulent particle acceleration in many astrophysical systems. However, we believe that it is reasonable to anticipate that $\alpha$ should approach an $L$-independent value for larger sizes, beyond those covered in our study. In particular, simulations of relativistic magnetic reconnection do provide evidence for an asymptotic regime in which power-law particle distributions become universal, but they only approach this regime for significantly larger systems than considered in our present work \citep{sironi_spitkovsky_2014, guo_etal_2014, werner_etal_2015}. One may therefore expect that a longer inertial range is needed to properly resolve all of the physical processes, including instabilities, responsible for turbulent particle acceleration. In addition, since only our largest simulations exhibit a convincing inertial range, finite-size corrections to turbulence statistics may feed into the particle acceleration properties. Finally, we note that in many astrophysical systems, additional processes such as particle escape (due to open boundaries) and radiative cooling can play an essential role in saturating the non-thermal distributions. The absence of such processes in our simulations prevents a proper steady-state distribution from being achieved, and makes it nontrivial to link the indices in our dynamically-evolving system to those in quasi-steady-state astrophysical systems. These important, subtle issues of convergence will await the next generation of turbulence simulations.

\acknowledgements

The authors thank N. Loureiro and J. TenBarge for helpful conversations. The authors also acknowledge support from NSF grant AST-1411879 and NASA Astrophysics Theory Program grant NNX16AB28G. DU gratefully acknowledges the hospitality of the Institute for Advanced Study and support from the Ambrose Monell Foundation. An award of computer time was provided by the Innovative and Novel Computational Impact on Theory and Experiment (INCITE) program. This research used resources of the Argonne Leadership Computing Facility, which is a DOE Office of Science User Facility supported under Contract DE-AC02-06CH11357.


\begin{thebibliography}{53}%
\makeatletter
\providecommand \@ifxundefined [1]{%
 \@ifx{#1\undefined}
}%
\providecommand \@ifnum [1]{%
 \ifnum #1\expandafter \@firstoftwo
 \else \expandafter \@secondoftwo
 \fi
}%
\providecommand \@ifx [1]{%
 \ifx #1\expandafter \@firstoftwo
 \else \expandafter \@secondoftwo
 \fi
}%
\providecommand \natexlab [1]{#1}%
\providecommand \enquote  [1]{``#1''}%
\providecommand \bibnamefont  [1]{#1}%
\providecommand \bibfnamefont [1]{#1}%
\providecommand \citenamefont [1]{#1}%
\providecommand \href@noop [0]{\@secondoftwo}%
\providecommand \href [0]{\begingroup \@sanitize@url \@href}%
\providecommand \@href[1]{\@@startlink{#1}\@@href}%
\providecommand \@@href[1]{\endgroup#1\@@endlink}%
\providecommand \@sanitize@url [0]{\catcode `\\12\catcode `\$12\catcode
  `\&12\catcode `\#12\catcode `\^12\catcode `\_12\catcode `\%12\relax}%
\providecommand \@@startlink[1]{}%
\providecommand \@@endlink[0]{}%
\providecommand \url  [0]{\begingroup\@sanitize@url \@url }%
\providecommand \@url [1]{\endgroup\@href {#1}{\urlprefix }}%
\providecommand \urlprefix  [0]{URL }%
\providecommand \Eprint [0]{\href }%
\providecommand \doibase [0]{http://dx.doi.org/}%
\providecommand \selectlanguage [0]{\@gobble}%
\providecommand \bibinfo  [0]{\@secondoftwo}%
\providecommand \bibfield  [0]{\@secondoftwo}%
\providecommand \translation [1]{[#1]}%
\providecommand \BibitemOpen [0]{}%
\providecommand \bibitemStop [0]{}%
\providecommand \bibitemNoStop [0]{.\EOS\space}%
\providecommand \EOS [0]{\spacefactor3000\relax}%
\providecommand \BibitemShut  [1]{\csname bibitem#1\endcsname}%
\let\auto@bib@innerbib\@empty
\bibitem [{\citenamefont {Sironi}\ and\ \citenamefont
  {Spitkovsky}(2014)}]{sironi_spitkovsky_2014}%
  \BibitemOpen
  \bibfield  {author} {\bibinfo {author} {\bibfnamefont {L.}~\bibnamefont
  {Sironi}}\ and\ \bibinfo {author} {\bibfnamefont {A.}~\bibnamefont
  {Spitkovsky}},\ }\href@noop {} {\bibfield  {journal} {\bibinfo  {journal}
  {The Astrophysical Journal Letters}\ }\textbf {\bibinfo {volume} {783}},\
  \bibinfo {pages} {L21} (\bibinfo {year} {2014})}\BibitemShut {NoStop}%
\bibitem [{\citenamefont {Guo}\ \emph {et~al.}(2014)\citenamefont {Guo},
  \citenamefont {Li}, \citenamefont {Daughton},\ and\ \citenamefont
  {Liu}}]{guo_etal_2014}%
  \BibitemOpen
  \bibfield  {author} {\bibinfo {author} {\bibfnamefont {F.}~\bibnamefont
  {Guo}}, \bibinfo {author} {\bibfnamefont {H.}~\bibnamefont {Li}}, \bibinfo
  {author} {\bibfnamefont {W.}~\bibnamefont {Daughton}}, \ and\ \bibinfo
  {author} {\bibfnamefont {Y.-H.}\ \bibnamefont {Liu}},\ }\href@noop {}
  {\bibfield  {journal} {\bibinfo  {journal} {Physical Review Letters}\
  }\textbf {\bibinfo {volume} {113}},\ \bibinfo {pages} {155005} (\bibinfo
  {year} {2014})}\BibitemShut {NoStop}%
\bibitem [{\citenamefont {Werner}\ \emph {et~al.}(2015)\citenamefont {Werner},
  \citenamefont {Uzdensky}, \citenamefont {Cerutti}, \citenamefont
  {Nalewajko},\ and\ \citenamefont {Begelman}}]{werner_etal_2015}%
  \BibitemOpen
  \bibfield  {author} {\bibinfo {author} {\bibfnamefont {G.}~\bibnamefont
  {Werner}}, \bibinfo {author} {\bibfnamefont {D.}~\bibnamefont {Uzdensky}},
  \bibinfo {author} {\bibfnamefont {B.}~\bibnamefont {Cerutti}}, \bibinfo
  {author} {\bibfnamefont {K.}~\bibnamefont {Nalewajko}}, \ and\ \bibinfo
  {author} {\bibfnamefont {M.}~\bibnamefont {Begelman}},\ }\href@noop {}
  {\bibfield  {journal} {\bibinfo  {journal} {The Astrophysical Journal
  Letters}\ }\textbf {\bibinfo {volume} {816}},\ \bibinfo {pages} {L8}
  (\bibinfo {year} {2015})}\BibitemShut {NoStop}%
\bibitem [{\citenamefont {Spitkovsky}(2008)}]{spitkovsky_2008}%
  \BibitemOpen
  \bibfield  {author} {\bibinfo {author} {\bibfnamefont {A.}~\bibnamefont
  {Spitkovsky}},\ }\href@noop {} {\bibfield  {journal} {\bibinfo  {journal}
  {The Astrophysical Journal Letters}\ }\textbf {\bibinfo {volume} {682}},\
  \bibinfo {pages} {L5} (\bibinfo {year} {2008})}\BibitemShut {NoStop}%
\bibitem [{\citenamefont {Sironi}\ and\ \citenamefont
  {Spitkovsky}(2010)}]{sironi_spitkovsky_2010}%
  \BibitemOpen
  \bibfield  {author} {\bibinfo {author} {\bibfnamefont {L.}~\bibnamefont
  {Sironi}}\ and\ \bibinfo {author} {\bibfnamefont {A.}~\bibnamefont
  {Spitkovsky}},\ }\href@noop {} {\bibfield  {journal} {\bibinfo  {journal}
  {The Astrophysical Journal}\ }\textbf {\bibinfo {volume} {726}},\ \bibinfo
  {pages} {75} (\bibinfo {year} {2010})}\BibitemShut {NoStop}%
\bibitem [{\citenamefont {Abdo}\ \emph {et~al.}(2011)\citenamefont {Abdo},
  \citenamefont {Ackermann}, \citenamefont {Ajello}, \citenamefont {Allafort},
  \citenamefont {Baldini}, \citenamefont {Ballet}, \citenamefont {Barbiellini},
  \citenamefont {Bastieri}, \citenamefont {Bechtol}, \citenamefont {Bellazzini}
  \emph {et~al.}}]{abdo_etal_2011}%
  \BibitemOpen
  \bibfield  {author} {\bibinfo {author} {\bibfnamefont {A.}~\bibnamefont
  {Abdo}}, \bibinfo {author} {\bibfnamefont {M.}~\bibnamefont {Ackermann}},
  \bibinfo {author} {\bibfnamefont {M.}~\bibnamefont {Ajello}}, \bibinfo
  {author} {\bibfnamefont {A.}~\bibnamefont {Allafort}}, \bibinfo {author}
  {\bibfnamefont {L.}~\bibnamefont {Baldini}}, \bibinfo {author} {\bibfnamefont
  {J.}~\bibnamefont {Ballet}}, \bibinfo {author} {\bibfnamefont
  {G.}~\bibnamefont {Barbiellini}}, \bibinfo {author} {\bibfnamefont
  {D.}~\bibnamefont {Bastieri}}, \bibinfo {author} {\bibfnamefont
  {K.}~\bibnamefont {Bechtol}}, \bibinfo {author} {\bibfnamefont
  {R.}~\bibnamefont {Bellazzini}},  \emph {et~al.},\ }\href@noop {} {\bibfield
  {journal} {\bibinfo  {journal} {Science}\ }\textbf {\bibinfo {volume}
  {331}},\ \bibinfo {pages} {739} (\bibinfo {year} {2011})}\BibitemShut
  {NoStop}%
\bibitem [{\citenamefont {Tavani}\ \emph {et~al.}(2011)\citenamefont {Tavani},
  \citenamefont {Bulgarelli}, \citenamefont {Vittorini}, \citenamefont
  {Pellizzoni}, \citenamefont {Striani}, \citenamefont {Caraveo}, \citenamefont
  {Weisskopf}, \citenamefont {Tennant}, \citenamefont {Pucella}, \citenamefont
  {Trois} \emph {et~al.}}]{tavani_etal_2011}%
  \BibitemOpen
  \bibfield  {author} {\bibinfo {author} {\bibfnamefont {M.}~\bibnamefont
  {Tavani}}, \bibinfo {author} {\bibfnamefont {A.}~\bibnamefont {Bulgarelli}},
  \bibinfo {author} {\bibfnamefont {V.}~\bibnamefont {Vittorini}}, \bibinfo
  {author} {\bibfnamefont {A.}~\bibnamefont {Pellizzoni}}, \bibinfo {author}
  {\bibfnamefont {E.}~\bibnamefont {Striani}}, \bibinfo {author} {\bibfnamefont
  {P.}~\bibnamefont {Caraveo}}, \bibinfo {author} {\bibfnamefont
  {M.}~\bibnamefont {Weisskopf}}, \bibinfo {author} {\bibfnamefont
  {A.}~\bibnamefont {Tennant}}, \bibinfo {author} {\bibfnamefont
  {G.}~\bibnamefont {Pucella}}, \bibinfo {author} {\bibfnamefont
  {A.}~\bibnamefont {Trois}},  \emph {et~al.},\ }\href@noop {} {\bibfield
  {journal} {\bibinfo  {journal} {Science}\ }\textbf {\bibinfo {volume}
  {331}},\ \bibinfo {pages} {736} (\bibinfo {year} {2011})}\BibitemShut
  {NoStop}%
\bibitem [{\citenamefont {Stone}\ \emph {et~al.}(1998)\citenamefont {Stone},
  \citenamefont {Ostriker},\ and\ \citenamefont {Gammie}}]{stone_etal_1998}%
  \BibitemOpen
  \bibfield  {author} {\bibinfo {author} {\bibfnamefont {J.~M.}\ \bibnamefont
  {Stone}}, \bibinfo {author} {\bibfnamefont {E.~C.}\ \bibnamefont {Ostriker}},
  \ and\ \bibinfo {author} {\bibfnamefont {C.~F.}\ \bibnamefont {Gammie}},\
  }\href@noop {} {\bibfield  {journal} {\bibinfo  {journal} {The Astrophysical
  Journal Letters}\ }\textbf {\bibinfo {volume} {508}},\ \bibinfo {pages} {L99}
  (\bibinfo {year} {1998})}\BibitemShut {NoStop}%
\bibitem [{\citenamefont {Servidio}\ \emph {et~al.}(2009)\citenamefont
  {Servidio}, \citenamefont {Matthaeus}, \citenamefont {Shay}, \citenamefont
  {Cassak},\ and\ \citenamefont {Dmitruk}}]{servidio_etal_2009}%
  \BibitemOpen
  \bibfield  {author} {\bibinfo {author} {\bibfnamefont {S.}~\bibnamefont
  {Servidio}}, \bibinfo {author} {\bibfnamefont {W.}~\bibnamefont {Matthaeus}},
  \bibinfo {author} {\bibfnamefont {M.}~\bibnamefont {Shay}}, \bibinfo {author}
  {\bibfnamefont {P.}~\bibnamefont {Cassak}}, \ and\ \bibinfo {author}
  {\bibfnamefont {P.}~\bibnamefont {Dmitruk}},\ }\href@noop {} {\bibfield
  {journal} {\bibinfo  {journal} {Physical Review Letters}\ }\textbf {\bibinfo
  {volume} {102}},\ \bibinfo {pages} {115003} (\bibinfo {year}
  {2009})}\BibitemShut {NoStop}%
\bibitem [{\citenamefont {Eyink}\ \emph {et~al.}(2013)\citenamefont {Eyink},
  \citenamefont {Vishniac}, \citenamefont {Lalescu}, \citenamefont {Aluie},
  \citenamefont {Kanov}, \citenamefont {B{\"u}rger}, \citenamefont {Burns},
  \citenamefont {Meneveau},\ and\ \citenamefont {Szalay}}]{eyink_etal_2013}%
  \BibitemOpen
  \bibfield  {author} {\bibinfo {author} {\bibfnamefont {G.}~\bibnamefont
  {Eyink}}, \bibinfo {author} {\bibfnamefont {E.}~\bibnamefont {Vishniac}},
  \bibinfo {author} {\bibfnamefont {C.}~\bibnamefont {Lalescu}}, \bibinfo
  {author} {\bibfnamefont {H.}~\bibnamefont {Aluie}}, \bibinfo {author}
  {\bibfnamefont {K.}~\bibnamefont {Kanov}}, \bibinfo {author} {\bibfnamefont
  {K.}~\bibnamefont {B{\"u}rger}}, \bibinfo {author} {\bibfnamefont
  {R.}~\bibnamefont {Burns}}, \bibinfo {author} {\bibfnamefont
  {C.}~\bibnamefont {Meneveau}}, \ and\ \bibinfo {author} {\bibfnamefont
  {A.}~\bibnamefont {Szalay}},\ }\href@noop {} {\bibfield  {journal} {\bibinfo
  {journal} {Nature}\ }\textbf {\bibinfo {volume} {497}},\ \bibinfo {pages}
  {466} (\bibinfo {year} {2013})}\BibitemShut {NoStop}%
\bibitem [{\citenamefont {Zhdankin}\ \emph {et~al.}(2013)\citenamefont
  {Zhdankin}, \citenamefont {Uzdensky}, \citenamefont {Perez},\ and\
  \citenamefont {Boldyrev}}]{zhdankin_etal_2013}%
  \BibitemOpen
  \bibfield  {author} {\bibinfo {author} {\bibfnamefont {V.}~\bibnamefont
  {Zhdankin}}, \bibinfo {author} {\bibfnamefont {D.~A.}\ \bibnamefont
  {Uzdensky}}, \bibinfo {author} {\bibfnamefont {J.~C.}\ \bibnamefont {Perez}},
  \ and\ \bibinfo {author} {\bibfnamefont {S.}~\bibnamefont {Boldyrev}},\
  }\href@noop {} {\bibfield  {journal} {\bibinfo  {journal} {The Astrophysical
  Journal}\ }\textbf {\bibinfo {volume} {771}},\ \bibinfo {eid} {124} (\bibinfo
  {year} {2013})}\BibitemShut {NoStop}%
\bibitem [{\citenamefont {Fermi}(1949)}]{fermi_1949}%
  \BibitemOpen
  \bibfield  {author} {\bibinfo {author} {\bibfnamefont {E.}~\bibnamefont
  {Fermi}},\ }\href@noop {} {\bibfield  {journal} {\bibinfo  {journal}
  {Physical Review}\ }\textbf {\bibinfo {volume} {75}},\ \bibinfo {pages}
  {1169} (\bibinfo {year} {1949})}\BibitemShut {NoStop}%
\bibitem [{\citenamefont {Blandford}\ and\ \citenamefont
  {Eichler}(1987)}]{blandford_eichler_1987}%
  \BibitemOpen
  \bibfield  {author} {\bibinfo {author} {\bibfnamefont {R.}~\bibnamefont
  {Blandford}}\ and\ \bibinfo {author} {\bibfnamefont {D.}~\bibnamefont
  {Eichler}},\ }\href@noop {} {\bibfield  {journal} {\bibinfo  {journal}
  {Physics Reports}\ }\textbf {\bibinfo {volume} {154}},\ \bibinfo {pages} {1}
  (\bibinfo {year} {1987})}\BibitemShut {NoStop}%
\bibitem [{\citenamefont {Cho}\ and\ \citenamefont
  {Lazarian}(2006)}]{cho_lazarian_2006}%
  \BibitemOpen
  \bibfield  {author} {\bibinfo {author} {\bibfnamefont {J.}~\bibnamefont
  {Cho}}\ and\ \bibinfo {author} {\bibfnamefont {A.}~\bibnamefont {Lazarian}},\
  }\href@noop {} {\bibfield  {journal} {\bibinfo  {journal} {The Astrophysical
  Journal}\ }\textbf {\bibinfo {volume} {638}},\ \bibinfo {pages} {811}
  (\bibinfo {year} {2006})}\BibitemShut {NoStop}%
\bibitem [{\citenamefont {Chandran}\ \emph {et~al.}(2010)\citenamefont
  {Chandran}, \citenamefont {Li}, \citenamefont {Rogers}, \citenamefont
  {Quataert},\ and\ \citenamefont {Germaschewski}}]{chandran_etal_2010}%
  \BibitemOpen
  \bibfield  {author} {\bibinfo {author} {\bibfnamefont {B.~D.}\ \bibnamefont
  {Chandran}}, \bibinfo {author} {\bibfnamefont {B.}~\bibnamefont {Li}},
  \bibinfo {author} {\bibfnamefont {B.~N.}\ \bibnamefont {Rogers}}, \bibinfo
  {author} {\bibfnamefont {E.}~\bibnamefont {Quataert}}, \ and\ \bibinfo
  {author} {\bibfnamefont {K.}~\bibnamefont {Germaschewski}},\ }\href@noop {}
  {\bibfield  {journal} {\bibinfo  {journal} {The Astrophysical Journal}\
  }\textbf {\bibinfo {volume} {720}},\ \bibinfo {pages} {503} (\bibinfo {year}
  {2010})}\BibitemShut {NoStop}%
\bibitem [{\citenamefont {Petrosian}(2012)}]{petrosian_2012}%
  \BibitemOpen
  \bibfield  {author} {\bibinfo {author} {\bibfnamefont {V.}~\bibnamefont
  {Petrosian}},\ }\href@noop {} {\bibfield  {journal} {\bibinfo  {journal}
  {Space science reviews}\ }\textbf {\bibinfo {volume} {173}},\ \bibinfo
  {pages} {535} (\bibinfo {year} {2012})}\BibitemShut {NoStop}%
\bibitem [{\citenamefont {Lazarian}\ \emph {et~al.}(2012)\citenamefont
  {Lazarian}, \citenamefont {Vlahos}, \citenamefont {Kowal}, \citenamefont
  {Yan}, \citenamefont {Beresnyak},\ and\ \citenamefont
  {Dal~Pino}}]{lazarian_etal_2012}%
  \BibitemOpen
  \bibfield  {author} {\bibinfo {author} {\bibfnamefont {A.}~\bibnamefont
  {Lazarian}}, \bibinfo {author} {\bibfnamefont {L.}~\bibnamefont {Vlahos}},
  \bibinfo {author} {\bibfnamefont {G.}~\bibnamefont {Kowal}}, \bibinfo
  {author} {\bibfnamefont {H.}~\bibnamefont {Yan}}, \bibinfo {author}
  {\bibfnamefont {A.}~\bibnamefont {Beresnyak}}, \ and\ \bibinfo {author}
  {\bibfnamefont {E.~d.~G.}\ \bibnamefont {Dal~Pino}},\ }\href@noop {}
  {\bibfield  {journal} {\bibinfo  {journal} {Space science reviews}\ }\textbf
  {\bibinfo {volume} {173}},\ \bibinfo {pages} {557} (\bibinfo {year}
  {2012})}\BibitemShut {NoStop}%
\bibitem [{\citenamefont {Beresnyak}\ and\ \citenamefont
  {Li}(2016)}]{beresnyak_li_2016}%
  \BibitemOpen
  \bibfield  {author} {\bibinfo {author} {\bibfnamefont {A.}~\bibnamefont
  {Beresnyak}}\ and\ \bibinfo {author} {\bibfnamefont {H.}~\bibnamefont {Li}},\
  }\href@noop {} {\bibfield  {journal} {\bibinfo  {journal} {The Astrophysical
  Journal}\ }\textbf {\bibinfo {volume} {819}},\ \bibinfo {pages} {90}
  (\bibinfo {year} {2016})}\BibitemShut {NoStop}%
\bibitem [{\citenamefont {Guo}\ \emph {et~al.}(2015)\citenamefont {Guo},
  \citenamefont {Liu}, \citenamefont {Daughton},\ and\ \citenamefont
  {Li}}]{guo_etal_2015}%
  \BibitemOpen
  \bibfield  {author} {\bibinfo {author} {\bibfnamefont {F.}~\bibnamefont
  {Guo}}, \bibinfo {author} {\bibfnamefont {Y.-H.}\ \bibnamefont {Liu}},
  \bibinfo {author} {\bibfnamefont {W.}~\bibnamefont {Daughton}}, \ and\
  \bibinfo {author} {\bibfnamefont {H.}~\bibnamefont {Li}},\ }\href@noop {}
  {\bibfield  {journal} {\bibinfo  {journal} {The Astrophysical Journal}\
  }\textbf {\bibinfo {volume} {806}},\ \bibinfo {pages} {167} (\bibinfo {year}
  {2015})}\BibitemShut {NoStop}%
\bibitem [{\citenamefont {Cho}(2005)}]{cho_2005}%
  \BibitemOpen
  \bibfield  {author} {\bibinfo {author} {\bibfnamefont {J.}~\bibnamefont
  {Cho}},\ }\href@noop {} {\bibfield  {journal} {\bibinfo  {journal} {The
  Astrophysical Journal}\ }\textbf {\bibinfo {volume} {621}},\ \bibinfo {pages}
  {324} (\bibinfo {year} {2005})}\BibitemShut {NoStop}%
\bibitem [{\citenamefont {Zrake}\ and\ \citenamefont
  {MacFadyen}(2011)}]{zrake_macfadyen_2011}%
  \BibitemOpen
  \bibfield  {author} {\bibinfo {author} {\bibfnamefont {J.}~\bibnamefont
  {Zrake}}\ and\ \bibinfo {author} {\bibfnamefont {A.~I.}\ \bibnamefont
  {MacFadyen}},\ }\href@noop {} {\bibfield  {journal} {\bibinfo  {journal} {The
  Astrophysical Journal}\ }\textbf {\bibinfo {volume} {744}},\ \bibinfo {pages}
  {32} (\bibinfo {year} {2011})}\BibitemShut {NoStop}%
\bibitem [{\citenamefont {Zrake}\ and\ \citenamefont
  {MacFadyen}(2012)}]{zrake_macfadyen_2012}%
  \BibitemOpen
  \bibfield  {author} {\bibinfo {author} {\bibfnamefont {J.}~\bibnamefont
  {Zrake}}\ and\ \bibinfo {author} {\bibfnamefont {A.~I.}\ \bibnamefont
  {MacFadyen}},\ }\href@noop {} {\bibfield  {journal} {\bibinfo  {journal} {The
  Astrophysical Journal Letters}\ }\textbf {\bibinfo {volume} {763}},\ \bibinfo
  {pages} {L12} (\bibinfo {year} {2012})}\BibitemShut {NoStop}%
\bibitem [{\citenamefont {Radice}\ and\ \citenamefont
  {Rezzolla}(2013)}]{radice_rezzolla_2013}%
  \BibitemOpen
  \bibfield  {author} {\bibinfo {author} {\bibfnamefont {D.}~\bibnamefont
  {Radice}}\ and\ \bibinfo {author} {\bibfnamefont {L.}~\bibnamefont
  {Rezzolla}},\ }\href@noop {} {\bibfield  {journal} {\bibinfo  {journal} {The
  Astrophysical Journal Letters}\ }\textbf {\bibinfo {volume} {766}},\ \bibinfo
  {pages} {L10} (\bibinfo {year} {2013})}\BibitemShut {NoStop}%
\bibitem [{\citenamefont {Cho}\ and\ \citenamefont
  {Lazarian}(2013)}]{cho_lazarian_2013}%
  \BibitemOpen
  \bibfield  {author} {\bibinfo {author} {\bibfnamefont {J.}~\bibnamefont
  {Cho}}\ and\ \bibinfo {author} {\bibfnamefont {A.}~\bibnamefont {Lazarian}},\
  }\href@noop {} {\bibfield  {journal} {\bibinfo  {journal} {The Astrophysical
  Journal}\ }\textbf {\bibinfo {volume} {780}},\ \bibinfo {pages} {30}
  (\bibinfo {year} {2013})}\BibitemShut {NoStop}%
\bibitem [{\citenamefont {Zrake}\ and\ \citenamefont
  {East}(2016)}]{zrake_east_2016}%
  \BibitemOpen
  \bibfield  {author} {\bibinfo {author} {\bibfnamefont {J.}~\bibnamefont
  {Zrake}}\ and\ \bibinfo {author} {\bibfnamefont {W.~E.}\ \bibnamefont
  {East}},\ }\href@noop {} {\bibfield  {journal} {\bibinfo  {journal} {The
  Astrophysical Journal}\ }\textbf {\bibinfo {volume} {817}},\ \bibinfo {pages}
  {89} (\bibinfo {year} {2016})}\BibitemShut {NoStop}%
\bibitem [{\citenamefont {Kowal}\ \emph {et~al.}(2012)\citenamefont {Kowal},
  \citenamefont {Dal~Pino},\ and\ \citenamefont {Lazarian}}]{kowal_etal_2012}%
  \BibitemOpen
  \bibfield  {author} {\bibinfo {author} {\bibfnamefont {G.}~\bibnamefont
  {Kowal}}, \bibinfo {author} {\bibfnamefont {E.~M. d.~G.}\ \bibnamefont
  {Dal~Pino}}, \ and\ \bibinfo {author} {\bibfnamefont {A.}~\bibnamefont
  {Lazarian}},\ }\href@noop {} {\bibfield  {journal} {\bibinfo  {journal}
  {Physical Review Letters}\ }\textbf {\bibinfo {volume} {108}},\ \bibinfo
  {pages} {241102} (\bibinfo {year} {2012})}\BibitemShut {NoStop}%
\bibitem [{\citenamefont {Lynn}\ \emph {et~al.}(2014)\citenamefont {Lynn},
  \citenamefont {Quataert}, \citenamefont {Chandran},\ and\ \citenamefont
  {Parrish}}]{lynn_etal_2014}%
  \BibitemOpen
  \bibfield  {author} {\bibinfo {author} {\bibfnamefont {J.~W.}\ \bibnamefont
  {Lynn}}, \bibinfo {author} {\bibfnamefont {E.}~\bibnamefont {Quataert}},
  \bibinfo {author} {\bibfnamefont {B.~D.}\ \bibnamefont {Chandran}}, \ and\
  \bibinfo {author} {\bibfnamefont {I.~J.}\ \bibnamefont {Parrish}},\
  }\href@noop {} {\bibfield  {journal} {\bibinfo  {journal} {The Astrophysical
  Journal}\ }\textbf {\bibinfo {volume} {791}},\ \bibinfo {pages} {71}
  (\bibinfo {year} {2014})}\BibitemShut {NoStop}%
\bibitem [{\citenamefont {{Makwana}}\ \emph {et~al.}(2015)\citenamefont
  {{Makwana}}, \citenamefont {{Zhdankin}}, \citenamefont {{Li}}, \citenamefont
  {{Daughton}},\ and\ \citenamefont {{Cattaneo}}}]{makwana_etal_2015}%
  \BibitemOpen
  \bibfield  {author} {\bibinfo {author} {\bibfnamefont {K.~D.}\ \bibnamefont
  {{Makwana}}}, \bibinfo {author} {\bibfnamefont {V.}~\bibnamefont
  {{Zhdankin}}}, \bibinfo {author} {\bibfnamefont {H.}~\bibnamefont {{Li}}},
  \bibinfo {author} {\bibfnamefont {W.}~\bibnamefont {{Daughton}}}, \ and\
  \bibinfo {author} {\bibfnamefont {F.}~\bibnamefont {{Cattaneo}}},\
  }\href@noop {} {\bibfield  {journal} {\bibinfo  {journal} {Physics of
  Plasmas}\ }\textbf {\bibinfo {volume} {22}},\ \bibinfo {eid} {042902}
  (\bibinfo {year} {2015})}\BibitemShut {NoStop}%
\bibitem [{\citenamefont {{Nalewajko}}\ \emph {et~al.}(2016)\citenamefont
  {{Nalewajko}}, \citenamefont {{Zrake}}, \citenamefont {{Yuan}}, \citenamefont
  {{East}},\ and\ \citenamefont {{Blandford}}}]{nalewajko_etal_2016}%
  \BibitemOpen
  \bibfield  {author} {\bibinfo {author} {\bibfnamefont {K.}~\bibnamefont
  {{Nalewajko}}}, \bibinfo {author} {\bibfnamefont {J.}~\bibnamefont
  {{Zrake}}}, \bibinfo {author} {\bibfnamefont {Y.}~\bibnamefont {{Yuan}}},
  \bibinfo {author} {\bibfnamefont {W.~E.}\ \bibnamefont {{East}}}, \ and\
  \bibinfo {author} {\bibfnamefont {R.~D.}\ \bibnamefont {{Blandford}}},\
  }\href {\doibase 10.3847/0004-637X/826/2/115} {\bibfield  {journal} {\bibinfo
   {journal} {The Astrophysical Journal}\ }\textbf {\bibinfo {volume} {826}},\
  \bibinfo {eid} {115} (\bibinfo {year} {2016})}\BibitemShut {NoStop}%
\bibitem [{\citenamefont {{Yuan}}\ \emph {et~al.}(2016)\citenamefont {{Yuan}},
  \citenamefont {{Nalewajko}}, \citenamefont {{Zrake}}, \citenamefont
  {{East}},\ and\ \citenamefont {{Blandford}}}]{yuan_etal_2016}%
  \BibitemOpen
  \bibfield  {author} {\bibinfo {author} {\bibfnamefont {Y.}~\bibnamefont
  {{Yuan}}}, \bibinfo {author} {\bibfnamefont {K.}~\bibnamefont {{Nalewajko}}},
  \bibinfo {author} {\bibfnamefont {J.}~\bibnamefont {{Zrake}}}, \bibinfo
  {author} {\bibfnamefont {W.~E.}\ \bibnamefont {{East}}}, \ and\ \bibinfo
  {author} {\bibfnamefont {R.~D.}\ \bibnamefont {{Blandford}}},\ }\href
  {\doibase 10.3847/0004-637X/828/2/92} {\bibfield  {journal} {\bibinfo
  {journal} {The Astrophysical Journal}\ }\textbf {\bibinfo {volume} {828}},\
  \bibinfo {eid} {92} (\bibinfo {year} {2016})}\BibitemShut {NoStop}%
\bibitem [{\citenamefont {{Makwana}}\ \emph {et~al.}(2016)\citenamefont
  {{Makwana}}, \citenamefont {{Li}}, \citenamefont {{Guo}},\ and\ \citenamefont
  {{Li}}}]{makwana_etal_2016_arxiv}%
  \BibitemOpen
  \bibfield  {author} {\bibinfo {author} {\bibfnamefont {K.}~\bibnamefont
  {{Makwana}}}, \bibinfo {author} {\bibfnamefont {H.}~\bibnamefont {{Li}}},
  \bibinfo {author} {\bibfnamefont {F.}~\bibnamefont {{Guo}}}, \ and\ \bibinfo
  {author} {\bibfnamefont {X.}~\bibnamefont {{Li}}},\ }\href@noop {} {\bibfield
   {journal} {\bibinfo  {journal} {ArXiv e-prints}\ } (\bibinfo {year}
  {2016})},\ \Eprint {http://arxiv.org/abs/1608.07829} {arXiv:1608.07829
  [physics.plasma-ph]} \BibitemShut {NoStop}%
\bibitem [{\citenamefont {Hoshino}(2013)}]{hoshino_2013}%
  \BibitemOpen
  \bibfield  {author} {\bibinfo {author} {\bibfnamefont {M.}~\bibnamefont
  {Hoshino}},\ }\href@noop {} {\bibfield  {journal} {\bibinfo  {journal} {The
  Astrophysical Journal}\ }\textbf {\bibinfo {volume} {773}},\ \bibinfo {pages}
  {118} (\bibinfo {year} {2013})}\BibitemShut {NoStop}%
\bibitem [{\citenamefont {Riquelme}\ \emph {et~al.}(2012)\citenamefont
  {Riquelme}, \citenamefont {Quataert}, \citenamefont {Sharma},\ and\
  \citenamefont {Spitkovsky}}]{riquelme_etal_2012}%
  \BibitemOpen
  \bibfield  {author} {\bibinfo {author} {\bibfnamefont {M.~A.}\ \bibnamefont
  {Riquelme}}, \bibinfo {author} {\bibfnamefont {E.}~\bibnamefont {Quataert}},
  \bibinfo {author} {\bibfnamefont {P.}~\bibnamefont {Sharma}}, \ and\ \bibinfo
  {author} {\bibfnamefont {A.}~\bibnamefont {Spitkovsky}},\ }\href@noop {}
  {\bibfield  {journal} {\bibinfo  {journal} {The Astrophysical Journal}\
  }\textbf {\bibinfo {volume} {755}},\ \bibinfo {pages} {50} (\bibinfo {year}
  {2012})}\BibitemShut {NoStop}%
\bibitem [{\citenamefont {Hoshino}(2015)}]{hoshino_2015}%
  \BibitemOpen
  \bibfield  {author} {\bibinfo {author} {\bibfnamefont {M.}~\bibnamefont
  {Hoshino}},\ }\href@noop {} {\bibfield  {journal} {\bibinfo  {journal}
  {Physical review letters}\ }\textbf {\bibinfo {volume} {114}},\ \bibinfo
  {pages} {061101} (\bibinfo {year} {2015})}\BibitemShut {NoStop}%
\bibitem [{\citenamefont {Cerutti}\ \emph {et~al.}(2013)\citenamefont
  {Cerutti}, \citenamefont {Werner}, \citenamefont {Uzdensky},\ and\
  \citenamefont {Begelman}}]{cerutti_etal_2013}%
  \BibitemOpen
  \bibfield  {author} {\bibinfo {author} {\bibfnamefont {B.}~\bibnamefont
  {Cerutti}}, \bibinfo {author} {\bibfnamefont {G.~R.}\ \bibnamefont {Werner}},
  \bibinfo {author} {\bibfnamefont {D.~A.}\ \bibnamefont {Uzdensky}}, \ and\
  \bibinfo {author} {\bibfnamefont {M.~C.}\ \bibnamefont {Begelman}},\
  }\href@noop {} {\bibfield  {journal} {\bibinfo  {journal} {The Astrophysical
  Journal}\ }\textbf {\bibinfo {volume} {770}},\ \bibinfo {pages} {147}
  (\bibinfo {year} {2013})}\BibitemShut {NoStop}%
\bibitem [{\citenamefont {Synge}(1957)}]{synge_1957}%
  \BibitemOpen
  \bibfield  {author} {\bibinfo {author} {\bibfnamefont {J.~L.}\ \bibnamefont
  {Synge}},\ }\href@noop {} {\emph {\bibinfo {title} {The relativistic gas}}},\
  Vol.~\bibinfo {volume} {32}\ (\bibinfo  {publisher} {North-Holland
  Amsterdam},\ \bibinfo {year} {1957})\BibitemShut {NoStop}%
\bibitem [{\citenamefont {Goldreich}\ and\ \citenamefont
  {Sridhar}(1995)}]{goldreich_sridhar_1995}%
  \BibitemOpen
  \bibfield  {author} {\bibinfo {author} {\bibfnamefont {P.}~\bibnamefont
  {Goldreich}}\ and\ \bibinfo {author} {\bibfnamefont {S.}~\bibnamefont
  {Sridhar}},\ }\href@noop {} {\bibfield  {journal} {\bibinfo  {journal} {The
  Astrophysical Journal}\ }\textbf {\bibinfo {volume} {438}},\ \bibinfo {pages}
  {763} (\bibinfo {year} {1995})}\BibitemShut {NoStop}%
\bibitem [{\citenamefont {TenBarge}\ \emph {et~al.}(2014)\citenamefont
  {TenBarge}, \citenamefont {Howes}, \citenamefont {Dorland},\ and\
  \citenamefont {Hammett}}]{tenbarge_etal_2014}%
  \BibitemOpen
  \bibfield  {author} {\bibinfo {author} {\bibfnamefont {J.}~\bibnamefont
  {TenBarge}}, \bibinfo {author} {\bibfnamefont {G.~G.}\ \bibnamefont {Howes}},
  \bibinfo {author} {\bibfnamefont {W.}~\bibnamefont {Dorland}}, \ and\
  \bibinfo {author} {\bibfnamefont {G.~W.}\ \bibnamefont {Hammett}},\
  }\href@noop {} {\bibfield  {journal} {\bibinfo  {journal} {Computer Physics
  Communications}\ }\textbf {\bibinfo {volume} {185}},\ \bibinfo {pages} {578}
  (\bibinfo {year} {2014})}\BibitemShut {NoStop}%
\bibitem [{\citenamefont {Sakai}\ and\ \citenamefont
  {Kawata}(1980)}]{sakai_kawata_1980}%
  \BibitemOpen
  \bibfield  {author} {\bibinfo {author} {\bibfnamefont {J.-i.}\ \bibnamefont
  {Sakai}}\ and\ \bibinfo {author} {\bibfnamefont {T.}~\bibnamefont {Kawata}},\
  }\href@noop {} {\bibfield  {journal} {\bibinfo  {journal} {Journal of the
  Physical Society of Japan}\ }\textbf {\bibinfo {volume} {49}},\ \bibinfo
  {pages} {747} (\bibinfo {year} {1980})}\BibitemShut {NoStop}%
\bibitem [{\citenamefont {Gedalin}(1993)}]{gedalin_1993}%
  \BibitemOpen
  \bibfield  {author} {\bibinfo {author} {\bibfnamefont {M.}~\bibnamefont
  {Gedalin}},\ }\href@noop {} {\bibfield  {journal} {\bibinfo  {journal}
  {Physical Review E}\ }\textbf {\bibinfo {volume} {47}},\ \bibinfo {pages}
  {4354} (\bibinfo {year} {1993})}\BibitemShut {NoStop}%
\bibitem [{Note1()}]{Note1}%
  \BibitemOpen
  \bibinfo {note} {Particle distributions are converged when $128$ or more
  particles per cell are used; see Supplementary Material at [URL will be
  inserted by publisher]}\BibitemShut {NoStop}%
\bibitem [{\citenamefont {Zhdankin}\ \emph {et~al.}(2016)\citenamefont
  {Zhdankin}, \citenamefont {Boldyrev},\ and\ \citenamefont
  {Uzdensky}}]{zhdankin_etal_2016b}%
  \BibitemOpen
  \bibfield  {author} {\bibinfo {author} {\bibfnamefont {V.}~\bibnamefont
  {Zhdankin}}, \bibinfo {author} {\bibfnamefont {S.}~\bibnamefont {Boldyrev}},
  \ and\ \bibinfo {author} {\bibfnamefont {D.~A.}\ \bibnamefont {Uzdensky}},\
  }\href@noop {} {\bibfield  {journal} {\bibinfo  {journal} {Physics of
  Plasmas}\ }\textbf {\bibinfo {volume} {23}},\ \bibinfo {pages} {055705}
  (\bibinfo {year} {2016})}\BibitemShut {NoStop}%
\bibitem [{\citenamefont {Wan}\ \emph {et~al.}(2016)\citenamefont {Wan},
  \citenamefont {Matthaeus}, \citenamefont {Roytershteyn}, \citenamefont
  {Parashar}, \citenamefont {Wu},\ and\ \citenamefont
  {Karimabadi}}]{wan_etal_2016}%
  \BibitemOpen
  \bibfield  {author} {\bibinfo {author} {\bibfnamefont {M.}~\bibnamefont
  {Wan}}, \bibinfo {author} {\bibfnamefont {W.}~\bibnamefont {Matthaeus}},
  \bibinfo {author} {\bibfnamefont {V.}~\bibnamefont {Roytershteyn}}, \bibinfo
  {author} {\bibfnamefont {T.}~\bibnamefont {Parashar}}, \bibinfo {author}
  {\bibfnamefont {P.}~\bibnamefont {Wu}}, \ and\ \bibinfo {author}
  {\bibfnamefont {H.}~\bibnamefont {Karimabadi}},\ }\href@noop {} {\bibfield
  {journal} {\bibinfo  {journal} {Physics of Plasmas (1994-present)}\ }\textbf
  {\bibinfo {volume} {23}},\ \bibinfo {pages} {042307} (\bibinfo {year}
  {2016})}\BibitemShut {NoStop}%
\bibitem [{\citenamefont {Kritsuk}\ \emph {et~al.}(2007)\citenamefont
  {Kritsuk}, \citenamefont {Norman}, \citenamefont {Padoan},\ and\
  \citenamefont {Wagner}}]{kritsuk_etal_2007}%
  \BibitemOpen
  \bibfield  {author} {\bibinfo {author} {\bibfnamefont {A.~G.}\ \bibnamefont
  {Kritsuk}}, \bibinfo {author} {\bibfnamefont {M.~L.}\ \bibnamefont {Norman}},
  \bibinfo {author} {\bibfnamefont {P.}~\bibnamefont {Padoan}}, \ and\ \bibinfo
  {author} {\bibfnamefont {R.}~\bibnamefont {Wagner}},\ }\href@noop {}
  {\bibfield  {journal} {\bibinfo  {journal} {The Astrophysical Journal}\
  }\textbf {\bibinfo {volume} {665}},\ \bibinfo {pages} {416} (\bibinfo {year}
  {2007})}\BibitemShut {NoStop}%
\bibitem [{\citenamefont {Federrath}\ \emph {et~al.}(2010)\citenamefont
  {Federrath}, \citenamefont {Roman-Duval}, \citenamefont {Klessen},
  \citenamefont {Schmidt},\ and\ \citenamefont
  {Mac~Low}}]{federrath_etal_2010}%
  \BibitemOpen
  \bibfield  {author} {\bibinfo {author} {\bibfnamefont {C.}~\bibnamefont
  {Federrath}}, \bibinfo {author} {\bibfnamefont {J.}~\bibnamefont
  {Roman-Duval}}, \bibinfo {author} {\bibfnamefont {R.}~\bibnamefont
  {Klessen}}, \bibinfo {author} {\bibfnamefont {W.}~\bibnamefont {Schmidt}}, \
  and\ \bibinfo {author} {\bibfnamefont {M.-M.}\ \bibnamefont {Mac~Low}},\
  }\href@noop {} {\bibfield  {journal} {\bibinfo  {journal} {Astronomy and
  Astrophysics}\ }\textbf {\bibinfo {volume} {512}},\ \bibinfo {pages} {A81}
  (\bibinfo {year} {2010})}\BibitemShut {NoStop}%
\bibitem [{\citenamefont {Schekochihin}\ \emph {et~al.}(2009)\citenamefont
  {Schekochihin}, \citenamefont {Cowley}, \citenamefont {Dorland},
  \citenamefont {Hammett}, \citenamefont {Howes}, \citenamefont {Quataert},\
  and\ \citenamefont {Tatsuno}}]{schekochihin_etal_2009}%
  \BibitemOpen
  \bibfield  {author} {\bibinfo {author} {\bibfnamefont {A.}~\bibnamefont
  {Schekochihin}}, \bibinfo {author} {\bibfnamefont {S.}~\bibnamefont
  {Cowley}}, \bibinfo {author} {\bibfnamefont {W.}~\bibnamefont {Dorland}},
  \bibinfo {author} {\bibfnamefont {G.}~\bibnamefont {Hammett}}, \bibinfo
  {author} {\bibfnamefont {G.}~\bibnamefont {Howes}}, \bibinfo {author}
  {\bibfnamefont {E.}~\bibnamefont {Quataert}}, \ and\ \bibinfo {author}
  {\bibfnamefont {T.}~\bibnamefont {Tatsuno}},\ }\href@noop {} {\bibfield
  {journal} {\bibinfo  {journal} {The Astrophysical Journal Supplement Series}\
  }\textbf {\bibinfo {volume} {182}},\ \bibinfo {pages} {310} (\bibinfo {year}
  {2009})}\BibitemShut {NoStop}%
\bibitem [{\citenamefont {Schoeffler}\ \emph {et~al.}(2014)\citenamefont
  {Schoeffler}, \citenamefont {Loureiro}, \citenamefont {Fonseca},\ and\
  \citenamefont {Silva}}]{schoeffler_etal_2014}%
  \BibitemOpen
  \bibfield  {author} {\bibinfo {author} {\bibfnamefont {K.}~\bibnamefont
  {Schoeffler}}, \bibinfo {author} {\bibfnamefont {N.}~\bibnamefont
  {Loureiro}}, \bibinfo {author} {\bibfnamefont {R.}~\bibnamefont {Fonseca}}, \
  and\ \bibinfo {author} {\bibfnamefont {L.}~\bibnamefont {Silva}},\
  }\href@noop {} {\bibfield  {journal} {\bibinfo  {journal} {Physical review
  letters}\ }\textbf {\bibinfo {volume} {112}},\ \bibinfo {pages} {175001}
  (\bibinfo {year} {2014})}\BibitemShut {NoStop}%
\bibitem [{\citenamefont {Thompson}\ and\ \citenamefont
  {Blaes}(1998)}]{thompson_blaes_1998}%
  \BibitemOpen
  \bibfield  {author} {\bibinfo {author} {\bibfnamefont {C.}~\bibnamefont
  {Thompson}}\ and\ \bibinfo {author} {\bibfnamefont {O.}~\bibnamefont
  {Blaes}},\ }\href@noop {} {\bibfield  {journal} {\bibinfo  {journal}
  {Physical Review D}\ }\textbf {\bibinfo {volume} {57}},\ \bibinfo {pages}
  {3219} (\bibinfo {year} {1998})}\BibitemShut {NoStop}%
\bibitem [{\citenamefont {Federrath}(2013)}]{federrath_2013}%
  \BibitemOpen
  \bibfield  {author} {\bibinfo {author} {\bibfnamefont {C.}~\bibnamefont
  {Federrath}},\ }\href@noop {} {\bibfield  {journal} {\bibinfo  {journal}
  {Monthly Notices of the Royal Astronomical Society}\ ,\ \bibinfo {pages}
  {stt1644}} (\bibinfo {year} {2013})}\BibitemShut {NoStop}%
\bibitem [{\citenamefont {Sahraoui}\ \emph {et~al.}(2009)\citenamefont
  {Sahraoui}, \citenamefont {Goldstein}, \citenamefont {Robert},\ and\
  \citenamefont {Khotyaintsev}}]{sahraoui_etal_2009}%
  \BibitemOpen
  \bibfield  {author} {\bibinfo {author} {\bibfnamefont {F.}~\bibnamefont
  {Sahraoui}}, \bibinfo {author} {\bibfnamefont {M.}~\bibnamefont {Goldstein}},
  \bibinfo {author} {\bibfnamefont {P.}~\bibnamefont {Robert}}, \ and\ \bibinfo
  {author} {\bibfnamefont {Y.~V.}\ \bibnamefont {Khotyaintsev}},\ }\href@noop
  {} {\bibfield  {journal} {\bibinfo  {journal} {Physical review letters}\
  }\textbf {\bibinfo {volume} {102}},\ \bibinfo {pages} {231102} (\bibinfo
  {year} {2009})}\BibitemShut {NoStop}%
\bibitem [{\citenamefont {Sahraoui}\ \emph {et~al.}(2013)\citenamefont
  {Sahraoui}, \citenamefont {Huang}, \citenamefont {Belmont}, \citenamefont
  {Goldstein}, \citenamefont {R{\'e}tino}, \citenamefont {Robert},\ and\
  \citenamefont {De~Patoul}}]{sahraoui_etal_2013}%
  \BibitemOpen
  \bibfield  {author} {\bibinfo {author} {\bibfnamefont {F.}~\bibnamefont
  {Sahraoui}}, \bibinfo {author} {\bibfnamefont {S.}~\bibnamefont {Huang}},
  \bibinfo {author} {\bibfnamefont {G.}~\bibnamefont {Belmont}}, \bibinfo
  {author} {\bibfnamefont {M.}~\bibnamefont {Goldstein}}, \bibinfo {author}
  {\bibfnamefont {A.}~\bibnamefont {R{\'e}tino}}, \bibinfo {author}
  {\bibfnamefont {P.}~\bibnamefont {Robert}}, \ and\ \bibinfo {author}
  {\bibfnamefont {J.}~\bibnamefont {De~Patoul}},\ }\href@noop {} {\bibfield
  {journal} {\bibinfo  {journal} {The Astrophysical Journal}\ }\textbf
  {\bibinfo {volume} {777}},\ \bibinfo {pages} {15} (\bibinfo {year}
  {2013})}\BibitemShut {NoStop}%
\bibitem [{\citenamefont {Alexandrova}\ \emph {et~al.}(2009)\citenamefont
  {Alexandrova}, \citenamefont {Saur}, \citenamefont {Lacombe}, \citenamefont
  {Mangeney}, \citenamefont {Mitchell}, \citenamefont {Schwartz},\ and\
  \citenamefont {Robert}}]{alexandrova_etal_2009}%
  \BibitemOpen
  \bibfield  {author} {\bibinfo {author} {\bibfnamefont {O.}~\bibnamefont
  {Alexandrova}}, \bibinfo {author} {\bibfnamefont {J.}~\bibnamefont {Saur}},
  \bibinfo {author} {\bibfnamefont {C.}~\bibnamefont {Lacombe}}, \bibinfo
  {author} {\bibfnamefont {A.}~\bibnamefont {Mangeney}}, \bibinfo {author}
  {\bibfnamefont {J.}~\bibnamefont {Mitchell}}, \bibinfo {author}
  {\bibfnamefont {S.~J.}\ \bibnamefont {Schwartz}}, \ and\ \bibinfo {author}
  {\bibfnamefont {P.}~\bibnamefont {Robert}},\ }\href@noop {} {\bibfield
  {journal} {\bibinfo  {journal} {Physical review letters}\ }\textbf {\bibinfo
  {volume} {103}},\ \bibinfo {pages} {165003} (\bibinfo {year}
  {2009})}\BibitemShut {NoStop}%
\bibitem [{\citenamefont {Alexandrova}\ \emph {et~al.}(2012)\citenamefont
  {Alexandrova}, \citenamefont {Lacombe}, \citenamefont {Mangeney},
  \citenamefont {Grappin},\ and\ \citenamefont
  {Maksimovic}}]{alexandrova_etal_2012}%
  \BibitemOpen
  \bibfield  {author} {\bibinfo {author} {\bibfnamefont {O.}~\bibnamefont
  {Alexandrova}}, \bibinfo {author} {\bibfnamefont {C.}~\bibnamefont
  {Lacombe}}, \bibinfo {author} {\bibfnamefont {A.}~\bibnamefont {Mangeney}},
  \bibinfo {author} {\bibfnamefont {R.}~\bibnamefont {Grappin}}, \ and\
  \bibinfo {author} {\bibfnamefont {M.}~\bibnamefont {Maksimovic}},\
  }\href@noop {} {\bibfield  {journal} {\bibinfo  {journal} {The Astrophysical
  Journal}\ }\textbf {\bibinfo {volume} {760}},\ \bibinfo {pages} {121}
  (\bibinfo {year} {2012})}\BibitemShut {NoStop}%
\end{thebibliography}

%

\end{document}